\documentclass[12pt,aps,prd,preprint,tightenlines,
amsmath,amssymb,showpacs,nofootinbib
]{revtex4}

\usepackage{amsmath}         
\usepackage{amssymb}         
\usepackage{amsfonts}        
\usepackage{graphicx}        

\usepackage{lipsum}        
\usepackage{color}         
\usepackage{enumerate}

\graphicspath{{figures/}}	

\renewcommand{\tilde}{\widetilde}   
\renewcommand{\email}[1]{\href{mailto:#1}{\texttt{#1}}}

\newcommand{\vect}[1]{\boldsymbol{#1}}		   

\let\oldenumerate\enumerate
\renewcommand{\enumerate}{
  \oldenumerate
  \setlength{\itemsep}{1pt}
  \setlength{\parskip}{0pt}
  \setlength{\parsep}{0pt}
}

\let\olditemize\itemize
\renewcommand{\itemize}{
  \olditemize
  \setlength{\itemsep}{1pt}
  \setlength{\parskip}{0pt}
  \setlength{\parsep}{0pt}
}

\newcommand{\PRE}[1]{{#1}}  

\newcommand{\K}{\text{K}}

\newcommand{\mev}{\text{MeV}}
\newcommand{\gev}{\text{GeV}}
\newcommand{\tev}{\text{TeV}}

\newcommand{\cm}{\text{cm}}
\newcommand{\m}{\text{m}}
\newcommand{\km}{\text{km}}
\newcommand{\g}{\text{g}}

\newcommand{\s}{\text{s}}

\renewcommand{\eqref}[1]{Eq.~(\ref{#1})}

\newcommand{\secref}[1]{Sec.~\ref{sec:#1}}

\newcommand{\figref}[1]{Fig.~\ref{fig:#1}}
\newcommand{\figsref}[2]{Figs.~\ref{fig:#1} and \ref{fig:#2}}

\newcommand{\tableref}[1]{Table~\ref{table:#1}}

\usepackage[colorlinks=true,citecolor=blue,linkcolor=blue,
urlcolor=blue,hypertexnames=false]{hyperref}

\begin{document}

\preprint{UCI-TR-2015-07}

\title{\PRE{\vspace*{1.5in}}
Dark Photons from the Center of the Earth: \\
Smoking-Gun Signals of Dark Matter 
\PRE{\vspace*{.5in}}}

\author{Jonathan L.~Feng\footnote{\email{jlf@uci.edu}}
}
\affiliation{Department of Physics and Astronomy, University of
  California, Irvine, California 92697, USA
\PRE{\vspace*{.4in}}
}

\author{Jordan Smolinsky\footnote{\email{jsmolins@uci.edu}}
}
\affiliation{Department of Physics and Astronomy, University of
  California, Irvine, California 92697, USA 
\PRE{\vspace*{.4in}}
}

\author{Philip Tanedo\footnote{\email{flip.tanedo@uci.edu}}
}
\affiliation{Department of Physics and Astronomy, University of
  California, Irvine, California 92697, USA
\PRE{\vspace*{.4in}}
}


\begin{abstract}
\PRE{\vspace*{.2in}} Dark matter may be charged under dark
electromagnetism with a dark photon that kinetically mixes with the
Standard Model photon.  In this framework, dark matter will collect at
the center of the Earth and annihilate into dark photons, which may
reach the surface of the Earth and decay into observable particles.
We determine the resulting signal rates, including Sommerfeld
enhancements, which play an important role in bringing the Earth's
dark matter population to their maximal, equilibrium value.  For dark
matter masses $m_X \sim 100~\gev - 10~\tev$, dark photon masses
$m_{A'} \sim \mev - \gev$, and kinetic mixing parameters
$\varepsilon \sim 10^{-10} - 10^{-8}$, the resulting electrons, muons,
photons, and hadrons that point back to the center of the Earth are a
smoking-gun signal of dark matter that may be detected by a variety of
experiments, including neutrino telescopes, such as IceCube, and
space-based cosmic ray detectors, such as Fermi-LAT and AMS.  We
determine the signal rates and characteristics, and show that large
and striking signals---such as parallel muon tracks---are possible 
in regions of the $(m_{A'}, \varepsilon)$ plane that are not probed 
by direct detection, accelerator experiments, or astrophysical observations.
\end{abstract}


\maketitle

\section{Introduction} 

Dark matter may live in a dark sector with its own forces. This
possibility has some nice features.  For example, the dark matter's
stability may be ensured not by some discrete parity imposed by hand,
but simply by its being the lightest fermion in the dark sector.  If
the dark sector contains an Abelian gauge symmetry, dark
electromagnetism, the dark photon and the Standard Model (SM) photon
will generically mix kinetically.  This mixing is of special
interest because it is one of the few ways for a dark sector to
interact with the known particles through a renormalizable interaction
and it is non-decoupling: a particle charged under both dark and
standard electromagnetism induces this interaction at loop-level, and
the effect is not suppressed for very heavy particles.  In this way,
this is a prototype for simplified dark matter models with light
mediators.  
The idea of a separate sector with its own photon~\cite{Kobzarev:1966qya, Okun:1982xi} and the further possibility of kinetic mixing between these two photons~\cite{Holdom:1985ag, Holdom:1986eq} were first explored long ago, 
and the myriad implications
for dark matter detection have recently attracted widespread
interest~\cite{Essig:2013lka,Foot:2014mia}.

In this framework, dark matter will collect in the center of the Earth
and annihilate to dark photons $XX \to A' A'$.  These dark photons may
then travel to near the surface of the Earth and decay to SM
particles, which may be detected in a variety of experiments, from
under-ice/underwater/underground experiments, such as the current
experiments IceCube, SuperK, and ANTARES, and future ones, such as
KM3NeT, IceCube II, DUNE, and HyperK, to space-based cosmic ray
detectors, such as the current experiments Fermi-LAT and AMS-02, and
future ones, such as CALET, ISS-CREAM, and others.  The resulting
signals of electrons, muons, photons, and hadrons that point back to
the center of the Earth are potentially striking signals of dark
matter.

The possibility of dark matter signals from the centers of large
astrophysical bodies was first proposed and investigated many years
ago~\cite{Freese:1985qw,Press:1985ug,Silk:1985ax,%
  Krauss:1985aaa,Griest:1986yu,Gaisser:1986ha,Gould:1987ju,%
  Gould:1987ir,1988ApJ...328..919G,Gould:1991hx}, 
  and there have been important advances
for the particular case of the Earth in recent
years~\cite{Damour:1998rh,Damour:1998vg,Gould:1999je,Lundberg:2004dn,%
  Peter:2009mi,Peter:2009mk,Peter:2009mm,Bruch:2009rp, Koushiappas:2009ee}.  Typically
these signals rely on annihilation to neutrinos, resulting in
single-particle signals with a continuum of energies.  In contrast,
dark photons decay into two charged particles, which may be seen at
the same time in a single experiment, and the total energy of these
charged particles is equal to the dark matter particle's mass,
producing potentially spectacular results.

\begin{figure}[tb]
\includegraphics[width=.99\linewidth]{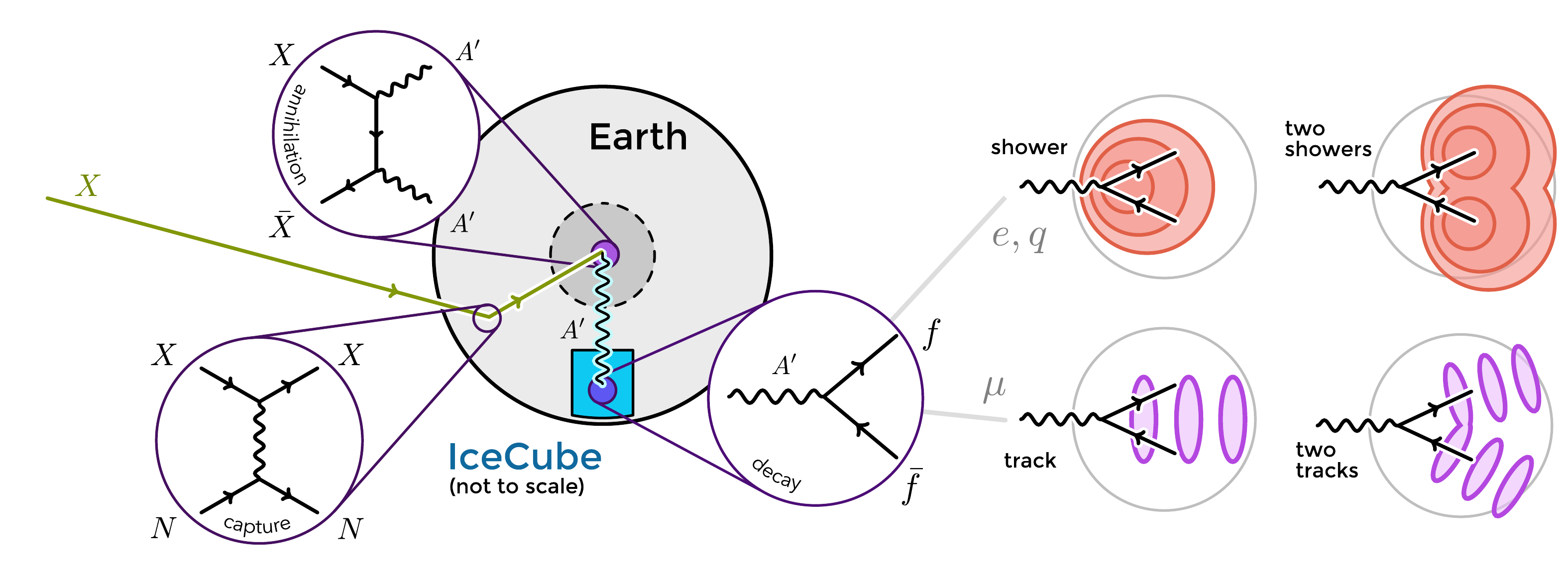}
\vspace*{-0.1in}
  \caption{Dark matter is captured by elastic $X N \to X N$ scattering
    off nuclei, collects in the center of the Earth, and annihilates
    to dark photons, $XX \to A' A'$.  These dark photons then travel
    to near the surface of the Earth and decay to SM particles, which
    may be detected by a variety of experiments, including neutrino
    telescopes and space-based cosmic ray detectors.  As an example,
    we show IceCube and various signatures there resulting from $A'$
    decays to electrons, muons, and hadrons. We discuss the
    possibility that double tracks (showers) may be resolved spatially
    (temporally) in the detector.}
  \label{fig:picture}
\vspace*{-0.1in}
\end{figure}

A schematic picture of this chain of events is given in
\figref{picture}. A number of processes must be evaluated to determine
the resulting signal.  For the specific case of dark photons, it is
tempting to simplify the analysis by making a number of assumptions.
For example, one may assume that the dark matter capture and
annihilation processes have reached equilibrium in the Earth and that
the capture cross section has some fixed value, such as the maximal
value consistent with current direct detection bounds.  Alternatively,
the calculations simplify immensely for dark matter masses large
compared to all relevant nuclear masses, $m_X\gg m_N$, or dark photon
masses $m_{A'}$ large compared to the characteristic momentum transfer
so that the interaction is point-like.  We show that none of these
assumptions are valid in the regions of parameter space of greatest
interest; the large $m_X$ approximation may lead to errors
 of an order
of magnitude for $m_X \approx 100~\gev$, and the large $m_{A'}$
approximation may also lead to mis-estimates of factors of a few for very light $m_{A'}\sim~\mev$. To
accurately determine the sensitivity of experiments to probe the
relevant parameter space, we carry out a general analysis, without
making these simplifying assumptions.  
An early exploration of dark matter accumulation on the Earth mediated by massless dark photons is Ref.~\cite{Holdom:1986eq}. For previous work exploring the case of massive dark photons,
see Ref.~\cite{Delaunay:2008pc} for the case of dark matter capturing in the Earth and annihilating into neutrinos, Refs.~\cite{Schuster:2009au,Schuster:2009fc,Meade:2009mu} for early work on celestial body capture of dark matter annihilating into dark photons, and in particular Ref.~\cite{Batell:2009zp} for a description of the general framework of annihilation to light mediators that highlights the specific case of solar capture and gamma ray signatures, which were later searched for by the Fermi-LAT collaboration~\cite{Ajello:2011dq}. Finally, recent work has highlighted the effect of self-capture~\cite{Albuquerque:2013xna,Chen:2015uha} and boosted dark matter~\cite{Berger:2014sqa}.

These results are timely for several reasons.  Dark photons have
attracted significant interest and are probed in many ways, including
direct detection experiments, accelerator and beam dump experiments,
and astrophysical observables~\cite{Essig:2013lka,Foot:2014mia}.  The
signals we discuss are detectable for dark photon masses $m_{A'} \sim
\mev - \gev$ and mixing parameters $\varepsilon \sim 10^{-10} -
10^{-8}$, an interesting and large region of parameter space that
includes territory that has not yet been probed.  These values of
$m_{A'}$ can also produce dark matter self-interactions that have been
suggested to solve small-scale structure
anomalies~\cite{Vogelsberger:2012ku,Rocha:2012jg,Peter:2012jh,Zavala:2012us,Tulin:2013teo}. The range of $\varepsilon$
values are naturally induced, for example, by degenerate
bi-fundamentals in grand unified theories~\cite{Collie:1998ty}.  It
was recently pointed out that combining kinetic mixings of this size
with the self-interacting models for small-scale structure can also
explain the excess of gamma rays from the galactic center recently
observed by Fermi-LAT~\cite{Kaplinghat:2015gha}.

At the same time, this work motivates a new class of searches for
current indirect detection experiments to discover dark matter.  At
present there are a number of landmark experiments, including those
mentioned above, that are transforming the field of indirect detection
with high precision measurements and increasingly large statistics.
In many cases, however, their sensitivities for dark matter searches
are clouded by uncertainties in astrophysical backgrounds.  The
signals we highlight here come from a specific direction (the center
of the Earth), cannot be mimicked by astrophysics and, in many cases,
are essentially background-free.  As a result, the processes discussed
here provide an opportunity for both current and future experiments to
detect a smoking-gun signal of dark matter.

\section{Dark Photons} 
\label{sec:darkphoton} 

We consider the simplest model of dark matter interacting through dark
photons.  The low-energy Lagrangian is
\begin{eqnarray}
{\cal L} &=& -\frac{1}{4} \tilde{F}_{\mu\nu} \tilde{F}^{\mu\nu}
- \frac{1}{4} \tilde{F}'_{\mu\nu} \tilde{F}'^{\mu\nu}
+ \frac{\epsilon}{2} \tilde{F}_{\mu\nu} \tilde{F}'^{\mu\nu}
- \frac{1}{2} m_{\tilde{A}'}^2 \tilde{A}'^2  \nonumber \\
&& + \sum_f \bar{f} (i \partial \! \! \! / - q_f e
\tilde{A} \! \! \! / - m_f) f
+ \bar{X} (i \partial \! \! \! / - g_X \tilde{A}' \! \! \! \! \! /
- m_X) X \ , 
\end{eqnarray}
where $X$ is the Dirac fermion dark matter, and $\tilde{A}$ and
$\tilde{A}'$ are the SM and dark sector gauge fields with field
strengths $\tilde{F}$ and $\tilde{F}'$ and fine-structure constants
$\alpha = e^2 / (4\pi)$ and $\alpha_X = g_X^2/(4\pi)$, respectively.
The sum is over SM fermions $f$ with SM electric charges $q_f$.  Dark
electromagnetism is broken and the mass $m_{\tilde{A}'}$ is generated
by some mechanism, such as the Higgs or Stueckelberg mechanisms, which
we assume otherwise plays a negligible role in the signals discussed
here. Note that the dark matter particles $X$ are stabilized not by
some {\em ad hoc} discrete parity symmetry or even by dark charge
conservation (which is broken), but by Lorentz symmetry, since $X$ is
the lightest fermion in the dark sector.

After diagonalizing the gauge kinetic and mass terms, the physical
states are the usual massless photon $A$, which does not couple to
$X$, and the dark photon $A'$ with mass $m_{A'} = m_{\tilde{A}'} /
\sqrt{1 - \epsilon^2}$, which couples both to $X$ and to SM fermions
with charge $\varepsilon q_f e$, where $\varepsilon \equiv \epsilon /
\sqrt{1 - \epsilon^2}$.  We take the independent parameters of the
theory to be
\begin{equation}
m_X \, , \ m_{A'}\, , \ \varepsilon \, , \ \alpha_X \ .
\label{eq:params}
\end{equation}
We typically fix $\alpha_X$ by requiring $X$ to saturate the observed
dark matter density through thermal freeze out, so $\alpha_X =
\alpha_X^{\text{th}} \simeq 0.035 (m_X / \tev)$.
Alternatively, the maximum allowed coupling is set by bounds on distortions to the cosmic microwave background~\cite{Adams:1998nr,Chen:2003gz,Padmanabhan:2005es}.
 Fitting the results from Ref.~\cite{Slatyer:2015jla}, we find $\alpha_X^\text{max} \simeq 0.17 (m_X / \tev)^{1.61}$.
With a choice of $\alpha_X$ the model
is completely determined by the first 3 parameters.

Dark photons decay to SM fermions with width
\begin{equation}
\Gamma ( A' \to f \bar{f}) = \frac{N_C \varepsilon^2 q_f^2 \alpha
  (m_{A'}^2 + 2 m_f^2)} {3 m_{A'}} \sqrt{1 - \frac{4 m_f^2
  }{m_{A'}^2} } \ ,
\end{equation}
where $N_C$ is the number of colors of fermion $f$.  The dark photons
we consider are produced from the annihilation of extremely
non-relativistic $X$ particles, and so have energy $m_X$. For $m_{A'}
\gg m_e$, the dark photon decay length is therefore
\begin{equation}
L = R_{\oplus} B_e 
\left( \frac{3.6 \! \times \! 10^{-9}}{\varepsilon} \right) ^2
\left( \frac{m_X / m_{A'}}{1000} \right)
\left( \frac{\gev}{m_{A'}} \right) ,
\end{equation}
where $R_{\oplus} \simeq 6370~\km$ is the radius of the Earth, and
$B_e \equiv B(A' \to e^+ e^-)$ is the branching fraction to electrons.
The $A'$ branching fractions can be determined from hadron production
at $e^+e^-$ colliders~\cite{Buschmann:2015awa}.  For $m_{A'} < 2
m_{\mu}$, $B_e = 100\%$.  As $m_{A'}$ increases above $2 m_{\mu}$, the
$A' \to \mu^+ \mu^-$ decay mode opens up rapidly, 
and $B_e$ drops to 50\% at $m_{A'} \sim 300~\mev$.  For $500~\mev \alt m_{A'} \alt 3~\gev$,
$B_e$ and $B_{\mu}$ are nearly identical and typically vary between
15\% and 40\%, with the rest made up by decays to hadrons, which also
produce photons and neutrinos from meson decays.  For $m_X$ at the
weak-scale and $m_{A'} \sim 100~\mev - \gev$, the requirement $L \sim
R_{\oplus}$ implies $\varepsilon \sim 10^{-10} - 10^{-8}$, and we will
see that this is indeed the range of kinetic mixing parameters that
gives the most promising signals.

\section{Dark Matter Accumulation in the Earth} 

Dark matter interacting through dark photons is captured and
annihilates at the center of the Earth.  The number $N_X$ of dark
matter particles in the Earth obeys the equation
\begin{align}
\frac{dN_X}{dt} = C_{\text{cap}} - C_{\text{ann}} N_X^2 \ ,
\label{NX:rate}
\end{align}
where $C_{\text{cap}}$ and $\Gamma_{\text{ann}} = \frac{1}{2}
C_{\text{ann}} N_X^2$ are the rates for the capture and annihilation
processes.  We ignore dark matter evaporation, which is negligible for
weak-scale dark matter masses~\cite{Griest:1986yu,Gaisser:1986ha}.  We
also ignore self-capture from dark matter--dark matter
self-interactions.  The impact of self-capture for the Earth is
suppressed by the fact that the escape velocity is low compared to
typical galactic dark matter velocities, and so typical dark matter
self-scatterings simply replace one captured dark matter particle with
another~\cite{Zentner:2009is}.

The solution to \eqref{NX:rate} is
\begin{align}
\Gamma_{\text{ann}} = \frac{1}{2} C_{\text{cap}} 
\tanh^2 \left( \frac{\tau_{\oplus}}{\tau} \right) \ ,
\label{annihilation:rate:from:capture}
\end{align}
where $\tau_{\oplus} \simeq 4.5~\text{Gyr}$ is the age of the Earth, and
$\tau = (C_{\text{cap}} C_{\text{ann}})^{-1/2}$ is the timescale for
the competing processes of capture and annihilation to reach
equilibrium.  To evaluate $\Gamma_{\text{ann}}$, we must therefore
evaluate both $C_\text{cap}$ and $C_\text{ann}$, which we now do in
turn.

\subsection{Dark Matter Capture}
\label{sec:capture}

Dark matter particles are captured when elastic scattering off nuclei
$N$ in the Earth reduces their velocity below the escape velocity.
The elastic scattering process $X N \to X N$ is mediated by
$t$-channel $A'$ exchange.  
The most relevant scattering targets, $N$, are iron and nickel; 
these and other elements are listed in \tableref{elements}.
In the center-of-mass frame, the cross
section is
\begin{eqnarray}
\left. 
\frac{d\sigma_N}{d\Omega} \right|_{\text{CM}} 
\! \! 
&=&
\frac{1}{(E_X + E_N)^2} 
\frac{2 \varepsilon^2 \alpha_X \alpha Z_N^2}{[2 \vect{p}^2 (1 -
\cos\theta_\text{CM}) +
    m_{A'}^2 ] ^2} 
    |F_N|^2
\nonumber 
\\
&\times & 
\! \! 
\left[ (E_X E_N + \vect{p}^2)^2 + (E_X E_N + \vect{p}^2
    \cos\theta_\text{CM})^2 - (m_X^2 + m_N^2) \vect{p}^2
(1-\cos\theta_\text{CM}) 
    \right] ,
\label{scattering}
\end{eqnarray}
where $E_N$, $Z_N$, $m_N$, and $F_N$ are the energy, electric charge,
mass, and nuclear form factor of target nucleus $N$, and $\vect{p}$ is
the center-of-mass 3-momentum of the dark matter.  Since the collision
is non-relativistic, $\vect{p}$ is negligible everywhere in
\eqref{scattering}, except possibly the denominator.  The cross
section may then be simplified to
\begin{equation}
\left. \frac{d\sigma_N}{d\Omega} \right|_{\text{CM}} \!
\approx 4 \varepsilon^2 \alpha_X \alpha Z_N^2
\frac{\mu_{N}^2}{(2 \vect{p}^2 (1 - \cos\theta_\text{CM}) + m_{A'}^2)^2}
|F_N|^2 \ .
\label{elastic:scattering:xsec}
\end{equation}
where $\mu_{N} \equiv m_N m_X / (m_N + m_X)$ is the reduced mass of the
$X$--$N$ system.

It is tempting to simplify the denominator by neglecting $\vect p$,
and reducing the $A'$ exchange to a contact interaction.  However, it
is not always true that $m_{A'}^2 \gg \vect{p}^2$ so that the latter
term may be neglected. The typical size of the momentum is $\vect{p}^2
\sim \mu_{N}^2 w^2$, where $w$ is the $X$ velocity in the lab
frame. Since capture typically occurs only for very small asymptotic
dark matter velocities, a reasonable choice would be $w=v_\oplus(r_N)
\approx 5\times 10^{-5}$, the escape velocity at the radius $r_N$ that
maximizes the radial number density $n_N(r)r^2$ of target nucleus $N$.
With these values, the contact interaction limit fails for $m_{A'}
\alt 3~\mev$. Rather than neglecting the momentum term altogether, a
slightly more sophisticated approach would be to make the substitution
$\vect{p}^2(1-\cos\theta_\text{CM}) \to \mu_{N}^2 w^2$.  In this work,
however, we keep the full $\vect{p}$ dependence in the propagator and
evaluate the capture rate numerically so that our results are valid
throughout parameter space.  We have confirmed that our results
reproduce those in the literature in the corners of parameter space
where simplifying assumptions are valid. For example, they match
Ref.~\cite{Baratella:2013fya} in the large-$m_{A'}$, point-like cross
section limit.

To determine capture rates, it is convenient to re-express the
differential cross section as a function of recoil energy $E_R =
\mu_{N}^2 w^2(1-\cos \theta_\text{CM})/m_N$ in the lab frame.  In the
non-relativistic limit the expression simplifies
to~\cite{Fornengo:2011sz}
\begin{equation}
\frac{d\sigma_N}{dE_R} \approx
8\pi \varepsilon^2 \alpha_X \alpha Z_N^2
\frac{m_N}{w^2(2m_N E_R + m_{A'}^2)^2}
\vert F_N \vert^2 \ .
\label{dsigN:dER:dark:photon}
\end{equation}
For $F_N$, we adopt the Helm form factor~\cite{Lewin:1995rx},
\begin{equation}
\vert F_N(E_R) \vert^2 = \exp\left[-E_R/E_N\right] \ ,
\label{eq:helm:F}
\end{equation}
where 
$E_N \equiv 0.114~\gev/A_N^{5/3}$ is a characteristic
energy scale for a target nucleus with atomic mass number $A_N$.

From this fundamental cross section we can determine the capture rate.
The differential rate of dark matter particles scattering off nuclei
with incident velocity $w$ at radius $r$ from the center of the Earth
and imparting recoil energy between $E_R$ and $E_R + d E_R$ is given
by
\begin{equation}
dC_{\text{cap}} = n_X \sum_N  n_N(r) \frac{d\sigma_N}{d E_R} w \,
f_{\oplus}(w, r) \, d^3 w \, d^3 r \, dE_R \ ,
\label{eq:differential:capture:rate}
\end{equation}
where $n_X=(0.3\text{ GeV}/\text{cm}^3)/m_X$ and $n_N(r)$ are the dark
matter and target nucleus number densities, respectively, and
$f_{\oplus} (w, r)$ is the velocity distribution of incident dark
matter at radius $r$, which is distorted from the free-space
Maxwell--Boltzmann distribution, $f(u)$, by the Earth's motion and
gravitational potential.  We follow the velocity notation introduced
by Gould~\cite{Gould:1987ju, Gould:1987ir,Gould:1991hx} where
$v_\oplus(r)$ is the escape velocity at radius $r$ and $u$ is the dark
matter velocity asymptotically far from the Earth.

The total capture rate is obtained by integrating
\eqref{eq:differential:capture:rate} over the region of parameter
space where the final state dark matter particle has energy less than
$m_X v_{\oplus}^2(r)/2$ and is thus gravitationally captured. The
escape velocity $v_{\oplus}(r)$ and number densities $n_N(r)$ are
determined straightforwardly from the density data enumerated in the
Preliminary Reference Earth Model~\cite{3599194}.  Following Edsj\"o
and Lundberg~\cite{Lundberg:2004dn}, the target number densities are
modeled by dividing the Earth into two layers, the core and the
mantle, with constant densities and elemental compositions given in
\tableref{elements}. The capture rate is then $C_\text{cap} = \sum_N
C_\text{cap}^N$, where the rate on target $N$ is
\begin{align}
C_{\text{cap}}^N 
&= 
n_X 
\int_0^{R_{\oplus}} dr \, 4\pi r^2 n_N(r) 
\int_0^{\infty} dw\, 4 \pi w^3 f_{\oplus}(w, r) 
\int_{E_\text{min}}^{E_\text{max}} dE_R \frac{d\sigma_N}{d E_R}  \;
\Theta(\Delta E) \ .
\label{eq:Ccap:N:general}
\end{align}
Here $\Theta(\Delta E) = \Theta(E_\text{max}-E_\text{min})$ imposes
the constraint that capture is kinematically possible by enforcing
that the minimum energy transfer, $E_\text{min}$, to gravitationally
capture the dark matter particle is smaller than the maximum recoil
energy kinematically allowed, $E_\text{max}$, corresponding to
$\cos\theta_\text{CM} = -1$.  Explicitly, these energies are
\begin{align}
E_\text{min} &= \frac 12 m_X \left[w^2 - v_{\oplus}^2(r)\right] &
E_\text{max} &= \frac{2\mu_{N}^2}{m_N}w^2 \ .
\end{align}

\begin{table}
 \begin{ruledtabular}
\begin{tabular}{lccc|lccc}
Element & Core MF & Mantle MF & $C_\text{cap}^N (\text{s}^{-1})$ &
Element & Core MF & Mantle MF & $C_\text{cap}^N (\text{s}^{-1})$
\\
\hline
Iron & 0.855 & 0.0626 & $9.43\times 10^{7}$ &
Chromium & 0.009 & 0.0026 & $8.98\times 10^{5}$ \\
Nickel & 0.052 & 0.00196 & $7.10\times 10^{6}$ &
Oxygen & 0 & 0.440 & $4.03\times 10^{5}$ \\
Silicon & 0.06 & 0.210 & $2.24\times 10^{6}$ &
Sulfur & 0.019 & 0.00025 & $2.41\times 10^{5}$ \\
Magnesium & 0 & 0.228 & $1.05\times 10^{6}$ &
Aluminum & 0 & 0.0235 & $1.62\times 10^{5}$ \\
Calcium & 0 & 0.0253 & $9.06\times 10^{5}$ &
Phosphorus & 0.002 & 0.00009 & $2.04\times 10^{4}$
\end{tabular}
 \caption{Mass fractions of the Earth's core and mantle for the
   elements most relevant for dark matter
   capture~\cite{holland_turekian_2003,Lundberg:2004dn}.  Also shown
   for each element is the capture rate $C^N_\text{cap}$ for $m_X =
   1~\tev$, $m_{A'}= 1~\gev$, $\varepsilon=10^{-8}$, and $\alpha_X=
   \alpha_X^{\text{th}} \simeq 0.035$ as a measure of the relevance of
   the nuclear target for dark matter capture.
\label{table:elements} }
 \end{ruledtabular}
 \end{table}
 
To make further progress, we must determine the distribution
$f_{\oplus}(w,r)$. A simple approach is to only include the effect of
the Earth's gravitational potential.  However, the Earth is within the
gravitational influence of the Sun, and one might expect the
acceleration of dark matter by the sun to suppress or eliminate the
capture of heavy dark matter particles by the Earth.  In 1991,
however, Gould argued that the interactions of dark matter with other
planets leads to diffusion of the dark matter population between bound
and unbound orbits and one could thus ignore the impact of the Sun's
gravitational field and treat the Earth in the ``free-space''
approximation to reasonable accuracy~\cite{1991ApJ...368..610G}.

More recently, however, this simple picture has been refined with both
potentially positive and negative implications.  In numerical work,
both Lundberg and Edsj\"o~\cite{Lundberg:2004dn} and
Peter~\cite{Peter:2009mi,Peter:2009mk,Peter:2009mm} have investigated
the influence of the Sun, Earth, Jupiter, and Venus in more detail,
tracking the possibility that the Earth's dark matter population is
suppressed when particles are kicked out of the solar system or
captured by the Sun.  For the case of supersymmetric WIMPs---that is,
dark matter with weak-scale mediators---they have found that these
effects can reduce the Earth's capture rate by one order of magnitude
or more, depending on the dark matter mass.  On the other hand,
simulations of galaxies with baryons have shown that dark matter
substructures may be pulled into the disk and create a significant and
relatively cold enhancement of the local dark matter density known as
a ``dark disk''~\cite{2009MNRAS.397...44R,2008MNRAS.389.1041R}.  For
the case of WIMP dark matter, this population may enhance indirect
detection signals from the Earth by up to three orders of
magnitude~\cite{Bruch:2009rp,Purcell:2009yp}. Note that the dark disk
has a velocity relative to our solar system that is $\sim 1/5$ that of
the ordinary dark matter halo~\cite{Peter:2009qj}. It is thus
plausible that the dark disk populates a region in phase space more amenable to Earth capture without significantly enhancing the
direct detection rate. 

As we show below, the dark photon case differs significantly from
WIMPs, because both the capture and annihilation rates are highly
velocity dependent.  One consequence of this is that $\tau_\oplus$ is
typically larger than $\tau$ in
\eqref{annihilation:rate:from:capture}, as opposed to the
conventional wisdom that the Earth has not reached its WIMP capacity.
It is therefore not possible to simply extrapolate the conclusions of
WIMP studies to the present framework. In addition, as our analysis is
valid for general dark matter and dark photon masses, inaccuracies in
the particle physics modeling are greatly reduced, and the
astrophysical uncertainties from dark disk and other effects are very
likely the dominant uncertainties entering the signal rate derivation.
These astrophysical phenomena are therefore clearly interesting and
important, but are beyond the scope of the present work.  Here, we use
the free-space approximation, not because it is the last word, but
because it provides a simple ``middle ground'' estimate, with both
suppressions and enhancements possible.

With the free-space assumption, we proceed as follows.  By energy
conservation, $w$ and $u$, the incident dark matter particle's
velocities in the Earth's and galactic frame, respectively, are
related by
\begin{equation}
w^2 = u^2 + v_{\oplus}^2(r) \ .
\label{eq:velocities:energy:conservation}
\end{equation}
The capture rate for a general $d\sigma_N/dE_R$ can then be rewritten
as
\begin{align}
C_\text{cap}^N &=
n_X 
\int_0^{R_\oplus}  dr \, 4\pi r^2 n_N(r)
\int_0^\infty du \, 4\pi u^2 f_\oplus(u) \frac{u^2+v_\oplus^2(r)}{u}
\int_{E_\text{min}}^{E_\text{max}}
dE_R \frac{d\sigma_N}{d E_R}  \;
\Theta(\Delta E) \ .
\label{DM:capture:rate:general:sigma}
\end{align}
Here $f_{\oplus}(u)$ is defined to be the angular-averaged and
annual-averaged velocity distribution in the rest frame of the
Earth~\cite{Vergados:1998ax},
\begin{equation}
f_{\oplus}(u) = 
\frac{1}{4} \int \! \! \! 
\int_{-1}^1 \! \! d \! \cos\theta \, 
d \! \cos\phi \, f\left[\left(u^2 + (V_{\odot}+V_{\oplus}\cos\gamma \cos\phi )^2 
+ 2u (V_{\odot} + V_{\oplus}\cos\gamma \cos\phi )\cos\theta \right)^{1/2}\right] ,
\end{equation}
where $V_{\odot} \simeq 220~\km/\s$ is the velocity of the Sun
relative to the galactic center, $V_{\oplus} \simeq 29.8~\km/\s$
is the velocity of the Earth relative to the Sun, and $\cos \gamma \approx 0.51$ is 
the angle of inclination of the Earth's orbital plane relative to the Sun's orbit.  
Many-body simulations and other considerations suggest a dark matter velocity
distribution in the galactic rest frame of the
form~\cite{Vogelsberger:2008qb,Fairbairn:2008gz,%
  Kuhlen:2009vh,Ling:2009eh,Lisanti:2010qx,Mao:2012hf}
\begin{equation}
f(u) = N_0
\left[\exp\left(\frac{v_{\text{gal}}^2 - u^2}{k u_0^2}\right) 
- 1 \right]^k \Theta(v_{\text{gal}} - u) \ ,
\end{equation}
where $N_0$ is a normalization constant, $v_{\text{gal}}$ is the
escape velocity from the galaxy, and the parameters describing the
distribution have typical values in the
ranges~\cite{Baratella:2013fya,Choi:2013eda}
\begin{align}
220~\km/\s < u_0 < 270~\km/\s	&&
450~\km/\s < v_{\text{gal}} < 650~\km/\s &&
1.5 < k <3.5\ .
\end{align}
We use the midpoint values of each of these, namely, $u_0 =
245~\km/\s$, $v_{\text{gal}} = 550~\km/\s$, and $k=2.5$. The truncated
Maxwell--Boltzmann distribution is recovered for $k=0$.

Upon inserting \eqref{dsigN:dER:dark:photon}, the $dE_R$ integral in
\eqref{DM:capture:rate:general:sigma} evaluates to
\begin{align}
	\int_{E_\text{min}}^{E_\text{max}} dE_R \frac{d\sigma_N}{dE_R} 
	&=
	\frac{2\pi \varepsilon^2 \alpha_X\alpha Z_N^2}{w^2 m_N E_N} 
e^{\frac{m_{A'}^2}{2m_NE_N}}
	\left[
	\frac{e^{-x_N}}{x_N} + \text{Ei}(-x_N)
	\right]^{x_N^\text{min}}_{x_N^\text{max}} \ ,
\end{align}
where we use the substitution variable $x_N$ and exponential integral
function~\cite{abramowitz1964handbook},
\begin{align}
	x_N &= \frac{2 m_N E_R + m_{A'}^2}{2m_N E_N}
& \text{Ei}(z) &\equiv -\int_{-z}^\infty dt \frac{e^{-t}}{t} \ .
\end{align}
The total rate is $C_\text{cap} = \sum C_\text{cap}^N =
32\pi^3 \varepsilon^2 \alpha_X \alpha n_X \sum_N Z_N^2 
(m_N E_N)^{-1}
\text{exp}\left(\frac{m_{A'}^2}{2m_NE_N}\right)
c_\text{cap}^N$, where
\begin{align}
c_\text{cap}^N &=
\int_0^{R_\oplus}  dr \, r^2 n_N(r)
\int_0^\infty du \, u f_\oplus(u) 
\Theta(\Delta x_N)
\left[
	\frac{e^{-x_N}}{x_N} + \text{Ei}(-x_N)
	\right]^{x_N^\text{min}}_{x_N^\text{max}} \ .
\label{eq:DM:capture:rate:full}
\end{align}
The capture rates $C_{\text{cap}}^N$ for various nuclei $N$ at a
representative point in parameter space are shown in
\tableref{elements}.

\subsection{Dark Matter Annihilation}

Once a dark matter particle is captured by the Earth, it repeatedly
re-scatters, drops to the center of the Earth, and eventually
thermalizes with the surrounding matter.  In the case of the Sun, the
dark matter thermalizes within the age of the Sun for $X$--proton
spin-independent scattering cross sections greater than $10^{-51}$,
$10^{-50}$, and $10^{-47}~\cm^2$ for $m_X = 100~\gev$, 1 TeV, and 10
TeV, respectively~\cite{Peter:2009mm}.  Similar studies of Earth
capture are not available.  However, we will find that, for the range
of parameters where an observable indirect signal is possible, the
direct detection $X$--proton cross sections are at least $\sigma_p \sim
10^{-48}~\cm^2$, corresponding to $X$--iron cross sections of
$\sigma_{\text{Fe}} \sim Z_{\text{Fe}}^2 (m_{\text{Fe}}/m_p)^2
\sigma_p \sim 10^{-42}~\cm^2$, many orders of magnitude larger than
required for thermalization in the Sun.  We therefore expect dark
matter to be thermalized in the Earth to an excellent approximation.

For thermalized dark matter, the annihilation rate parameter
$C_{\text{ann}}$ is~\cite{Baratella:2013fya}
\begin{equation}
C_{\text{ann}} = \langle \sigma_{\text{ann}} v \rangle 
\left[ \frac{G_N m_X \rho_{\oplus}}{3 T_{\oplus}} \right] ^{3/2} \ ,
\end{equation}
where $\rho_{\oplus} \approx 13~\g/\cm^3$ and $T_{\oplus} \approx
5700~\K$ are the matter density and temperature at the center of the
Earth, respectively, $\sigma_{\text{ann}}$ is the cross section for
$XX \to A' A'$, and $v$ is the relative velocity of the interacting
particles, which is double the velocity of either interacting particle
in the center-of-mass frame.

The thermally-averaged cross section is
\begin{equation}
\langle \sigma_{\text{ann}} v \rangle 
= (\sigma_{\text{ann}} v)_{\text{tree}} \, \langle S_S \rangle \ ,
\end{equation}
where
\begin{equation}
(\sigma_{\text{ann}} v)_{\text{tree}} 
= \frac{\pi \alpha_X^2}{m_X^2} 
\frac{[1 - m_{A'}^2/m_X^2]^{3/2}} 
{[1-m_{A'}^2 / (2 m_X^2) ]^2} 
\end{equation}
is the tree-level cross section~\cite{Liu:2014cma}, and $\langle S_S
\rangle$ is the thermal average of the $S$-wave Sommerfeld enhancement
factor.  This Sommerfeld enhancement factor~\cite{Sommerfeld:1931} has
been determined with various degrees of refinement.  An analytic
expression that includes the resonance behavior present for non-zero 
$m_{A'}$ can be derived by approximating the Yukawa  potential by the
Hulth\'en potential~\cite{Cassel:2009wt,Slatyer:2009vg,Feng:2010zp}.
The resulting Sommerfeld factor is
\begin{equation}
	S_S = \frac{\pi}{a} \frac{\sinh(2\pi a c)}{\cosh(2\pi a c) -
          \cos(2\pi\sqrt{c-a^2c^2})} \ ,
\label{sommerfeld}
\end{equation}
where $a = v/(2\alpha_X)$ and $c=6\alpha_X m_X/(\pi^2 m_{A'})$. The
thermal average is, then,
\begin{equation}
\langle S_S \rangle = \int \frac{d^3v}{(2\pi v_0^2)^{3/2}} \,
e^{-\frac{1}{2} v^2 / v_0^2} \, S_S \ ,	
\label{eq:sommerfeld:full}
\end{equation}
where $v_0 = \sqrt{2T_{\oplus}/m_X}$.

\subsection{Equilibrium Time Scales}

\begin{figure}[t]
\includegraphics[width=0.41\linewidth]{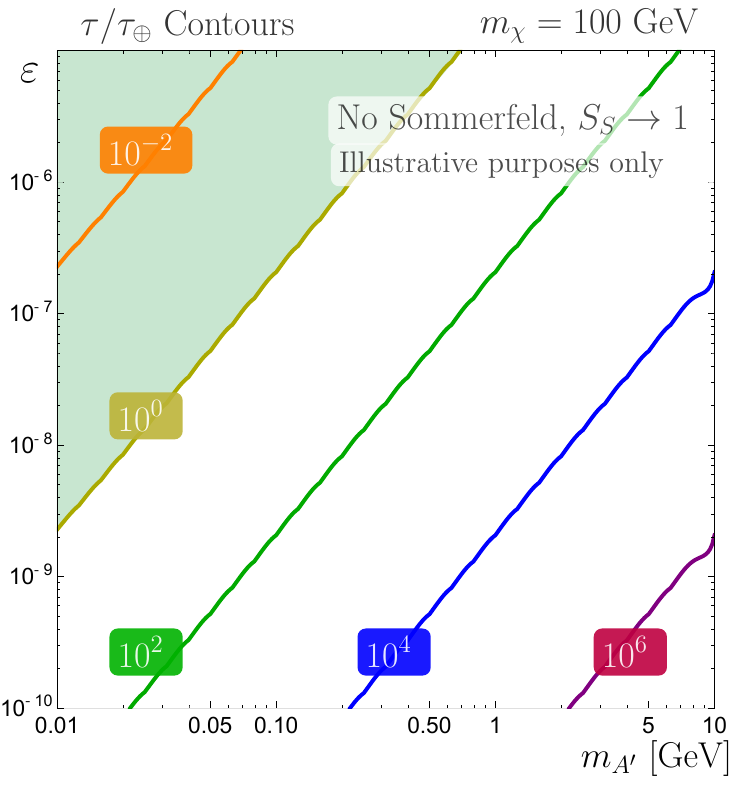} \qquad
\includegraphics[width=0.41\linewidth]{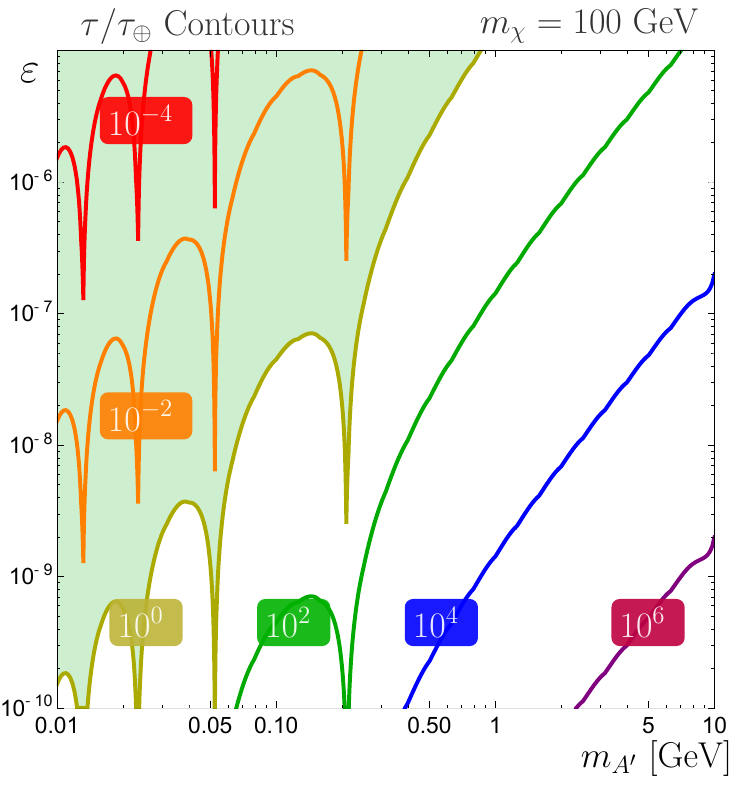} \\
\includegraphics[width=0.41\linewidth]{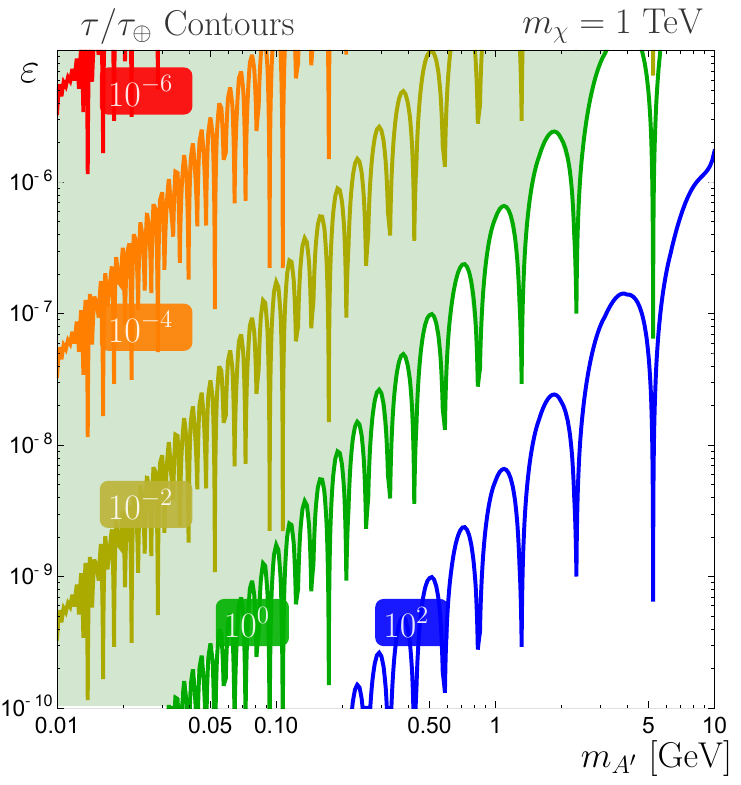} \qquad
\includegraphics[width=0.41\linewidth]{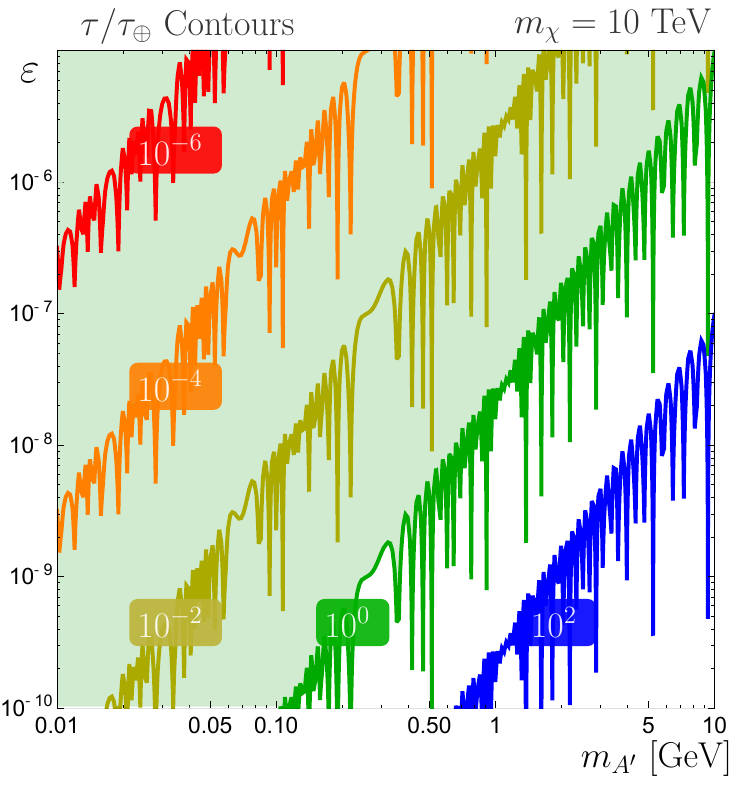} 
\vspace*{-0.1in}
\caption{Contours of constant $\tau / \tau_{\oplus}$, the equilibrium
  timescale in units of the Earth's lifetime, in the $(m_{A'},
  \varepsilon)$ plane for $m_X = 100~\gev$ without (top left) and with
  (top right) the Sommerfeld enhancement factor, $m_X = 1~\tev$ with
  the Sommerfeld factor (bottom left), and $m_X = 10~\tev$ with the
  Sommerfeld factor (bottom right).  The dark sector fine-structure
  constant $\alpha_X$ is set by requiring $\Omega_X \simeq 0.23$. In
  the green shaded regions, the Earth's lifetime is greater than the
  equilibrium timescale, $\tau_\oplus > \tau$.  }
  \label{fig:tauearth}
\vspace*{-0.1in}
\end{figure}

In \figref{tauearth} we present results for the equilibrium timescale
$\tau = (C_{\text{cap}} C_{\text{ann}})^{-1/2}$ in the $(m_{A'},
\varepsilon)$ plane for $m_X = 100~\gev$ without Sommerfeld enhancement,
and $m_X = 100~\gev, 1~\tev, 10~\tev$ with Sommerfeld enhancement.
The dark coupling $\alpha_X$ is fixed by the thermal relic density.
For $m_{A'} \ll m_X$, the parametric dependence of $\tau$ on
$\varepsilon$ and $m_{A'}$ enters dominantly through $C_{\text{ann}}$ and
is $\tau \sim C_{\text{ann}}^{-1/2} \sim m_{A'}^2 / \varepsilon$.  This
can be seen in the baseline values of the contours in
\figref{tauearth}.  The bumps in the contours reflect the resonance
structure of the Sommerfeld enhancement factor $\langle S_S \rangle$.

In the shaded green (upper left) parts of the figures, the Earth's dark matter
population has reached its maximal (equilibrium) value, and so the
annihilation rate is essentially determined by the capture rate, with
$\Gamma_{\text{ann}} \approx \frac{1}{2} C_{\text{cap}}$.  As one
moves down and to the right, however, the equilibrium timescale grows,
and the population is eventually not maximal.  We will see that when
the population is not at its maximal value, the signal quickly becomes
undetectable.  

The Sommerfeld enhancement plays an essential role in reducing the
equilibrium timescale and making the signal detectable in large
regions of the $(m_{A'}, \varepsilon)$ plane.  For capture, the
typical velocity that enters has an irreducible contribution from the
gravitational potential that accelerates dark matter as it falls into
the Earth. Capture interactions, therefore, occur at the typical
escape velocity in the Earth's core, $v_\text{esc} \approx 5.0 \times
10^{-5}$.  However, after the dark matter particles are captured, they
sink to the core, and come into thermal equilibrium with the normal
matter.  As a result, the population of dark matter particles at the
center of the Earth is even colder, with relative velocities $v_0
\approx 1.0\times 10^{-6} \, [\tev/m_X]^{1/2}$.  In the $m_{A'} \ll
\alpha_X m_X$ limit, the Sommerfeld factor of \eqref{sommerfeld}
becomes
\begin{equation}
S_0 = \frac{2 \pi \, \alpha_X / v} {1 - e^{- 2 \pi \alpha_X / v}} \ .
\label{sommerfeldsimple}
\end{equation}
For thermal relics, $S_0$ is therefore typically $\sim 2\pi \alpha_X /
v \sim 10^4 - 10^6$.  Sommerfeld enhancement therefore reduces the
equilibrium timescale by factors of $\sim 100$ for $m_X\sim 100~\gev$, 
as can be seen in \figref{tauearth} by comparing the top right and top left panels. This reduction on $\tau$ goes to $\sim 1000$ for $m_X\sim 10~\tev$.
The Sommerfeld factor therefore plays an essential role in boosting the
current Earth's dark matter population and the dark matter
signal~\cite{Delaunay:2008pc}.

\section{Signal Rates and Characteristics}

After dark matter accumulates in the center of the Earth and
annihilates to dark photons, the dark photons propagate outwards with
essentially no interactions with matter.  The characteristic radius of
the thermalized dark matter distribution in the 
Earth is~\cite{Baratella:2013fya}
\begin{equation}
r_X = \left(\frac{3 T_{\oplus}}
{2 \pi G_N \rho_{\oplus} m_X}\right)^{1/2} 
\approx 150~\km \sqrt{\frac{\tev}{m_X}} \ .
\label{eq:DM:radius:earth}
\end{equation}
An observer at the surface of the Earth or in low Earth orbit
therefore sees the majority of dark matter annihilations take place
within $1.3^{\circ} \sqrt{\tev/m_X}$ of straight down.

The dark photons are highly boosted with energy $m_X$.  In the
decay $A' \to f \bar{f}$, the characteristic angle between the
direction of a parent $A'$ and its decay products in the Earth's rest
frame is
\begin{equation}
\theta \sim \tan^{-1} \left( 
\frac{m_{A'}^2 - 4 m_f^2}{m_X^2 - m_{A'}^2} \right)^{1/2} 
\approx \frac{\sqrt{m_{A'}^2 - 4m_f^2}}{m_X} \ ,
\label{eq:opening:angle}
\end{equation}
assuming $m_{A'} \ll m_X$.  Much larger opening angles are possible,
however, as discussed in detail in the Appendix.

The indirect detection signal is therefore two highly collimated
leptons or jets that point back to the center of the Earth within a
few degrees.  As we will discuss, in some cases, the two leptons or
jets may be simultaneously detected, and possibly even seen as two
different particles, in contrast to the standard neutrino-based
indirect detection signals, where there is only one primary particle.
In any case, the signal of high-energy particles from the center of
the Earth distinguishes the signal from all possible astrophysical
backgrounds, potentially providing a smoking-gun signal of dark
matter if the event rates are large enough.

We now determine the event rates and characteristics for two classes
of experiments: under-ice/underground/underwater detectors,
represented by IceCube, and space-based experiments, represented by
Fermi-LAT and AMS-02.

\subsection{IceCube}

Dark photons may be detected in IceCube if they decay in IceCube or
just below it.  Decays $A' \to e^+ e^-, q\bar{q}$ will be seen as
showers, and for $m_{A'} \agt 300~\mev$, typically 15\% -- 40\% of the
decays will be to muons~\cite{Buschmann:2015awa} and be seen as
tracks.  The number of dark photon decays that can be detected by
IceCube is
\begin{equation}
N_{\text{sig}} = 2 \, \Gamma_{\text{ann}} \, 
\frac{A_{\text{eff}}}{4 \pi R_{\oplus}^2} \,
\epsilon_{\text{decay}} \, T \ ,
\end{equation}
where the factor of 2 results from the fact that each annihilation
produces two dark photons, $A_{\text{eff}}$ is the effective area of
detector,
\begin{equation}
\epsilon_{\text{decay}} = e^{-R_{\oplus}/L} - e^{-(R_{\oplus}+D)/L}
\label{eq:decay:efficiency}
\end{equation}
is the probability that the dark photon decays after traveling a
distance between $R_{\oplus}$ and $R_{\oplus}+D$, where $D$ is the
effective depth of the detector, and $T$ is the live time of the
experiment.

To very roughly estimate the detection rates for IceCube, we expect
that for $m_X \sim 1~\tev$, all dark photons that decay within the
instrumented volume of IceCube are detected, and so we take
$A_{\text{eff}} \approx 1~\km^2$ and $D \approx 1~\km$. For lighter
dark matter, say, $m_X \sim 100~\gev$, the decay products may be lost
between the photomultiplier strings of IceCube.  But these should be
seen with high efficiency in DeepCore~\cite{Collaboration:2011ym}, the
subset of IceCube with finer string spacings and lower threshold, and
so we also present results for the instrumented volume of DeepCore,
with $A_{\text{eff}} \approx 0.067~\km^2$ and $D \approx 0.55~\km$.

\begin{figure}[h] 
\hspace*{-.5cm}
\includegraphics[width=0.45\linewidth]{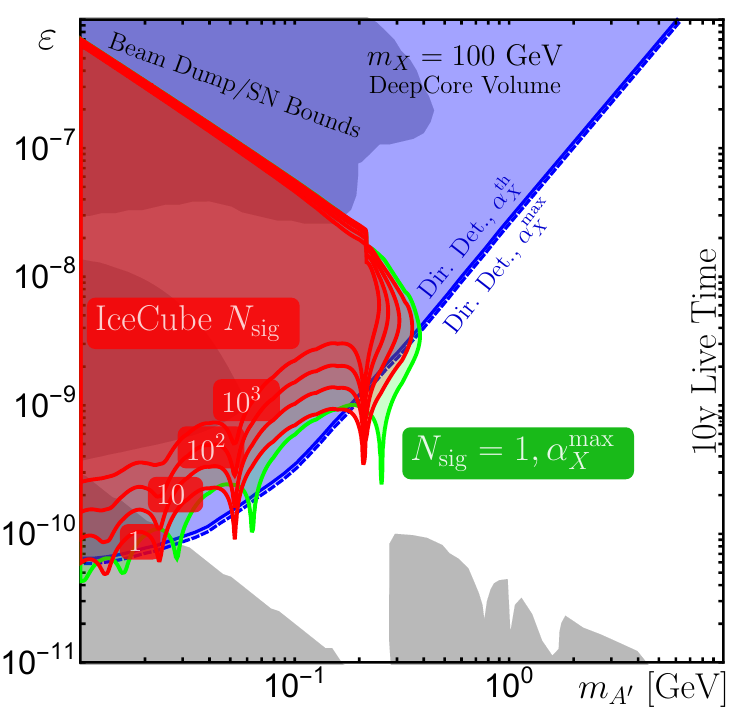} \qquad
\includegraphics[width=0.45\linewidth]{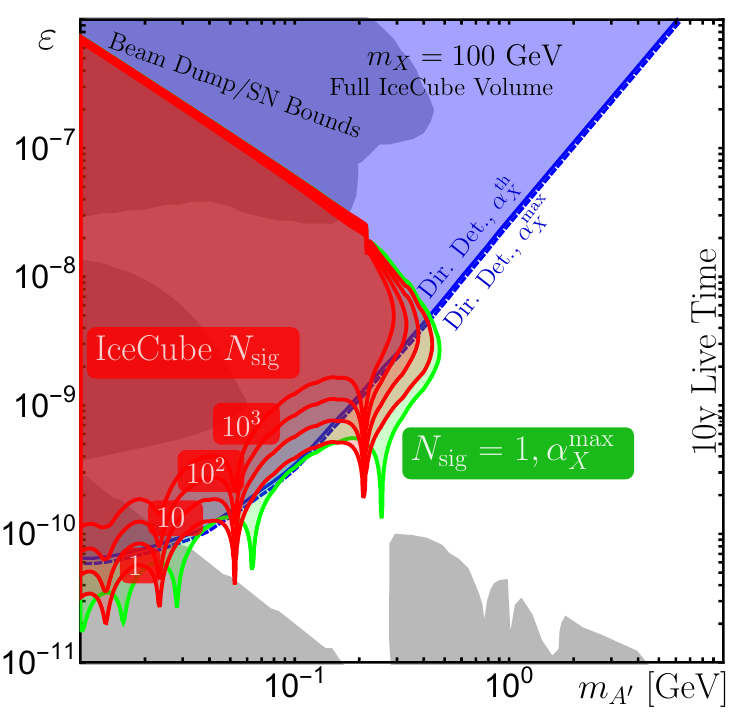} \\
\includegraphics[width=0.45\linewidth]{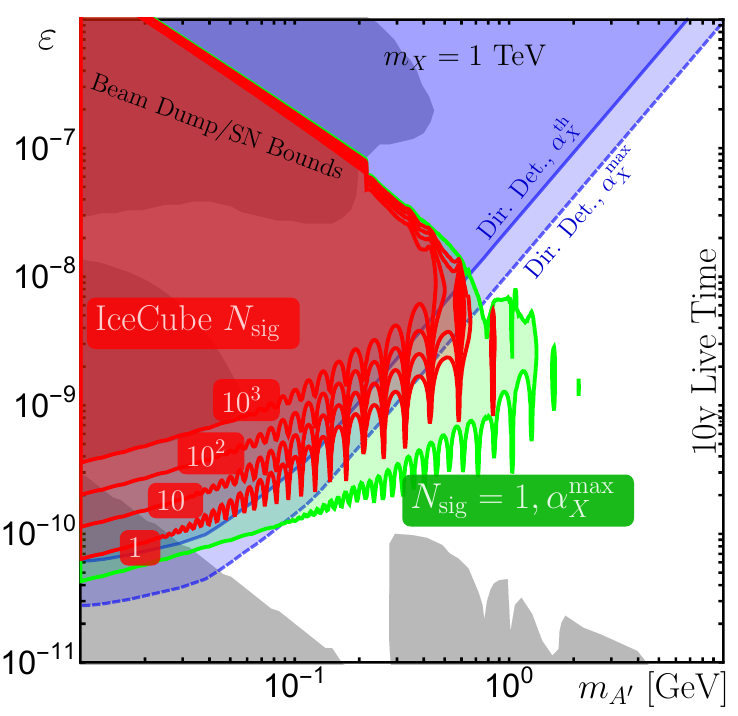} \qquad
\includegraphics[width=0.45\linewidth]{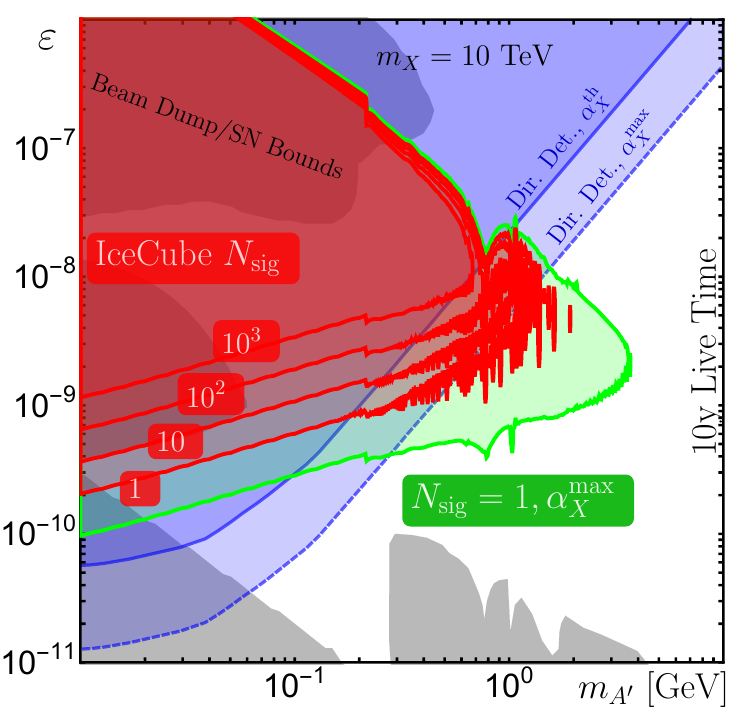}
\vspace*{-0.1in} 
  \caption{%
  \textbf{Red:} IceCube event rates for $T=10$ years live
    time in the $(m_{A'}, \varepsilon)$ plane for $m_X = 100~\gev$ in
    DeepCore (top left) and IceCube (top right), $m_X = 1~\tev$ in
    IceCube (bottom left), and $m_X = 10~\tev$ in IceCube (bottom
    right).  The dark sector fine-structure constant is set to the value $\alpha_X^\text{th}$ which realizes $\Omega_X \simeq 0.23$.  
    \textbf{Green:} Single event reach for the maximal dark fine-structure constant $\alpha_X^\text{max}$ allowed by cosmic microwave background distortion bounds~\cite{Slatyer:2015jla}. 
    \textbf{Blue:} current bounds from
    direct detection~\cite{DelNobile:2015uua}.  
    \textbf{Gray:} regions
    probed by beam dump and supernovae
    constraints~\cite{Dent:2012mx, Dreiner:2013mua, Essig:2013lka, Kazanas:2014mca, Rrapaj:2015wgs}.
    }
  \label{fig:resultsIceCube}
\vspace*{-0.1in}
\end{figure}

In \figref{resultsIceCube} we present the number of signal events for
$m_X = 100~\gev$, 1 TeV, and 10 TeV in the $(m_{A'}, \varepsilon)$ plane.
The bumpy features and closed contours are real physical features
resulting from Sommerfeld enhancement resonances.  Also shown are the
regions of parameter space disfavored by existing bounds on
dark-photon-mediated $X N \to X N$ scattering from direct detection
experiments, such as PANDAX-II~\cite{DelNobile:2015uua,Cui:2017nnn}, and $X$-independent
bounds on dark photons from beam
dump experiments and supernovae~\cite{Dent:2012mx, Dreiner:2013mua, Essig:2013lka, Kazanas:2014mca, Rrapaj:2015wgs,Mahoney:2017jqk}. We use the recently updated supernova cooling bounds in Ref.~\cite{Mahoney:2017jqk}.

\begin{figure}
\hspace*{-.5cm}
\includegraphics[width=0.45\linewidth]{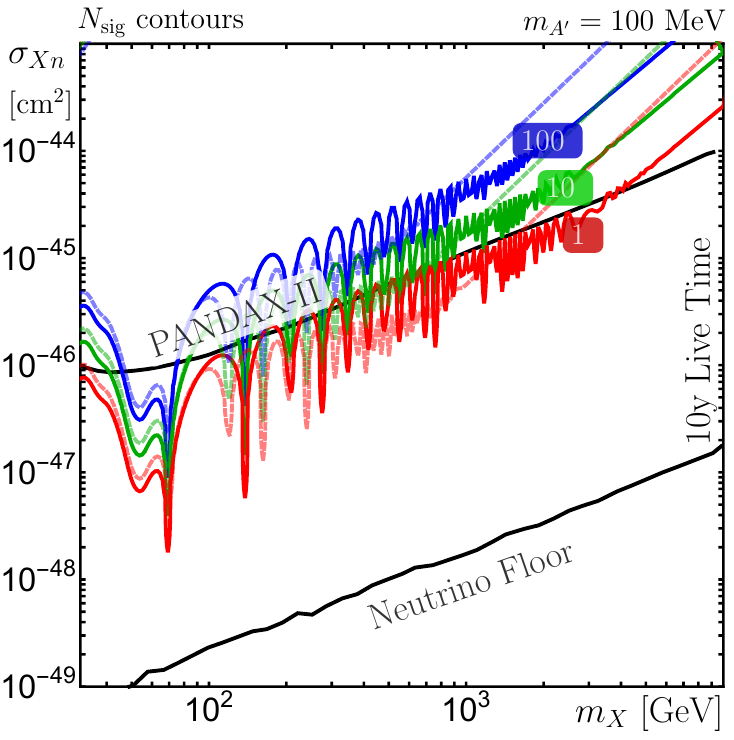} \qquad
\includegraphics[width=0.45\linewidth]{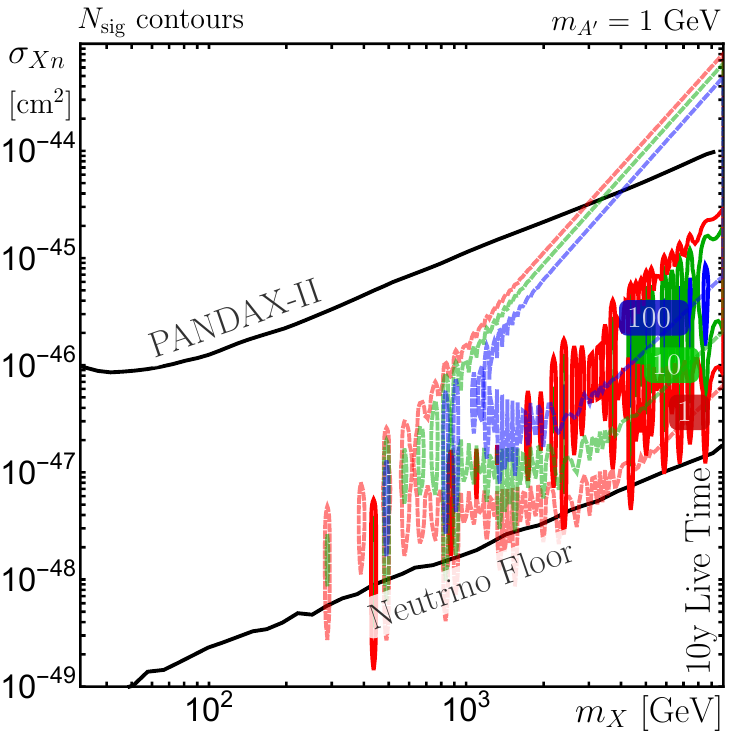} 
\vspace*{-0.1in}
  \caption{Comparison of indirect and direct detection sensitivities
    in the $(m_X, \sigma_{Xn})$ plane for $m_{A'} = 100~\mev$ (left) and 1
    GeV (right). 
    The direct detection bounds are from the LUX 
    collaboration~\cite{Akerib:2013tjd}. In this regime the interaction is effective point-like in contrast to
    the low $m_{A'}$ region~\cite{Kaplinghat:2013yxa, An:2014twa,
      DelNobile:2015uua} in \figref{resultsIceCube},
    where the direct detection bounds become independent of $m_{A'}$
    for low $m_{A'}$.  Also shown is the `neutrino floor,' where
    coherent neutrino scattering affects direct detection
    experiments~\cite{Billard:2013qya}; the dashed line is an
    extrapolation.
    }
\label{fig:direct}
\vspace*{-0.1in}
\end{figure}

We see that the indirect detection signal discussed here probes
regions of parameter space that are so far inaccessible by other
methods.  As anticipated in \secref{darkphoton}, the indirect
detection signal is largest for $\varepsilon \sim 10^{-10} - 10^{-8}$,
where the $A'$ decay length is comparable to $R_{\oplus}$.  For $m_X
\sim 10~\tev$ and $\varepsilon \sim 10^{-8}$, for example,
$N_{\text{sig}} \sim 1,000$ events over 10 years are possible in
regions of parameter space that are otherwise currently 
viable.  IceCube has collected roughly 7 years of data already, and so
detailed analyses will either exclude large new regions of the
$(m_{A'}, \varepsilon)$ parameter space or discover dark matter.

For $m_X \sim 100~\gev$, the indirect and direct detection
sensitivities are comparable for $\alpha_X$ between $\alpha_X^\text{th}$ and $\alpha_X^{\text{max}}$. 
The indirect and direct detection sensitivities
are shown in the conventional $(m_X, \sigma_{Xn})$ plane in
\figref{direct}, where $\sigma_{Xn}$ is the spin-independent $X$--nucleon
cross section: $\mu_{T}^2 A_T^2 \sigma_{Xn} = \mu_{n}^2 \sigma_{XT}$ with $T=\text{Xe}$.  The indirect detection signal is suppressed for both
 large $\sigma_{Xn}$ (large $\varepsilon$, dark photons decay too
soon) and small $\sigma_{Xn}$ (small $\varepsilon$, dark matter capture
is too slow and the captive population does not equilibrate).  Of
course, the large $\sigma_{Xn}$ are already excluded by direct detection
experiments.  Focusing on the small $\sigma_{Xn}$ region, for $m_X >
100~\gev$, the indirect detection signals probe cross sections as much
as three orders of magnitude below the current bounds from direct
detection experiments, such as XENON and LUX.

Since $\sigma_{Xn} \sim \alpha_X \varepsilon^2$, a given $\sigma_{Xn}$ corresponds to a larger value of $\varepsilon$ when assuming the thermal $\alpha_X^\text{th}$ versus maximal $\alpha_X^\text{max}$ dark sector coupling. For this reason, the dashed $\alpha_X^\text{max}$ curves on the left-hand plot in \figref{direct} are sometimes above the solid $\alpha_X^\text{th}$ curves on the $(m_X,\sigma_{Xn})$ plane. 
This is because when going from $\alpha_X^\text{th}$ to $\alpha_X^\text{max}$, the additional $\varepsilon$ reach gained in indirect detection experiments is less than that in direct detection experiments.
The reason for this is straightforward: the lower bound on the IceCube reach is set by the condition that the $\tanh^2 (\tau_\oplus/\tau)$ in Eq.~(\ref{annihilation:rate:from:capture}) is `saturated' near unity, i.e.\ that dark matter capture and annihilation are in equilibrium. This is why the lower contours of \figref{resultsIceCube} display the same resonances as Fig.~\ref{fig:tauearth}. Since this condition is set by the geometric mean of the capture and annihilation rates, it scales differently from direct detection experiments which has the same parametric dependence as the capture rate. 
%


The detector's effective area $A_{\text{eff}}$ and depth $D$ are, of
course, dependent on the energy and type of the dark photon decay
products, and a more detailed study of detector response is required
to estimate these more accurately.  This is beyond the scope of the
present work, but we note here some basic considerations.  Muons
with energies $E_{\mu} \sim 100~\gev - \tev$ lose energy primarily
through ionization and travel a distance
\begin{equation}
L_{\mu} = \frac{1}{\rho\beta} \ln
\left[ \frac{\alpha + \beta E_{\mu}}{\alpha + \beta E_{\text{th}}}
    \right]
\label{muonL}
\end{equation}
before their energy drops below a threshold energy $E_{\text{th}}$,
where $\rho = 1~\g/\cm^3$, $\alpha \simeq 2.0~\mev~\cm^2/\g$, and
$\beta \simeq 4.2 \times 10^{-6}~\cm^2/\g$~\cite{Agashe:2014kda}. For
$E_{\mu} = 1~\tev$ and $E_{\text{th}} = 50~\gev$, on average muons
travel a distance $L_{\mu} = 2.5~\km$.  Dark photons that decay to
muons a km or two below IceCube may therefore be detected in IceCube,
and so the effective depth of IceCube is a bit larger than 1 km.  For
$m_X \sim 10~\tev$, the effective depth is larger still, although less
than a naive application of \eqref{muonL} would indicate, as such high
energy muons lose energy primarily through radiative processes.  At
TeV energies, the experimental angular resolution for muon tracks is
less than a degree, providing a powerful handle to reduce background.
For the case of showers from electrons or hadrons, 
the angular resolution is worse, but still
sufficient to identify showers as up-going to within tens of degrees.
In addition, because dark photons decay completely to visible
particles, contained events are mono-energetic, with the total energy
equal to $m_X$.  The angle and energy distributions of tracks and
showers are therefore completely different from astrophysical sources,
and provide powerful handles for differentiating signal from
background.

\begin{figure}[t] 
\includegraphics[width=0.32\linewidth]{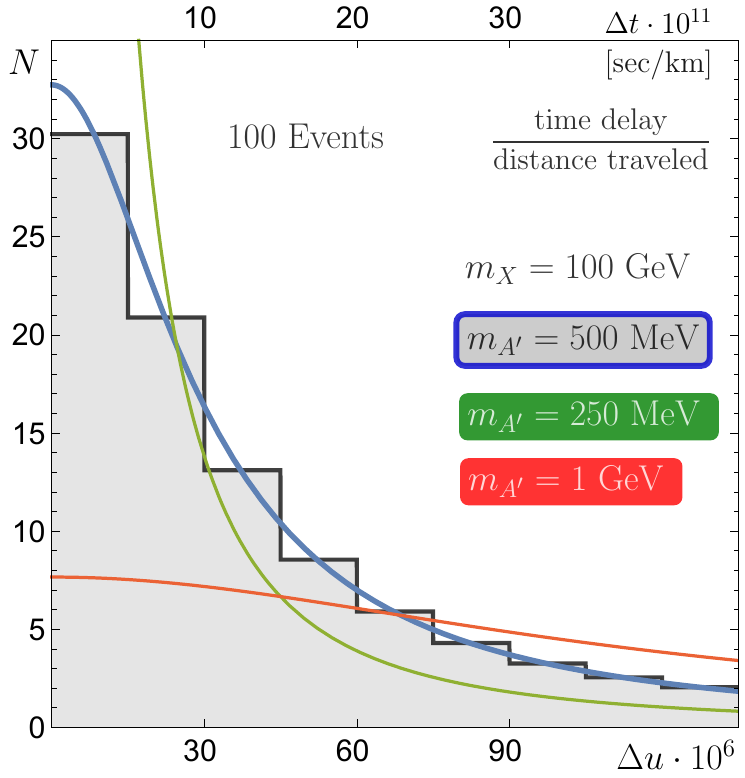} 
\includegraphics[width=0.32\linewidth]{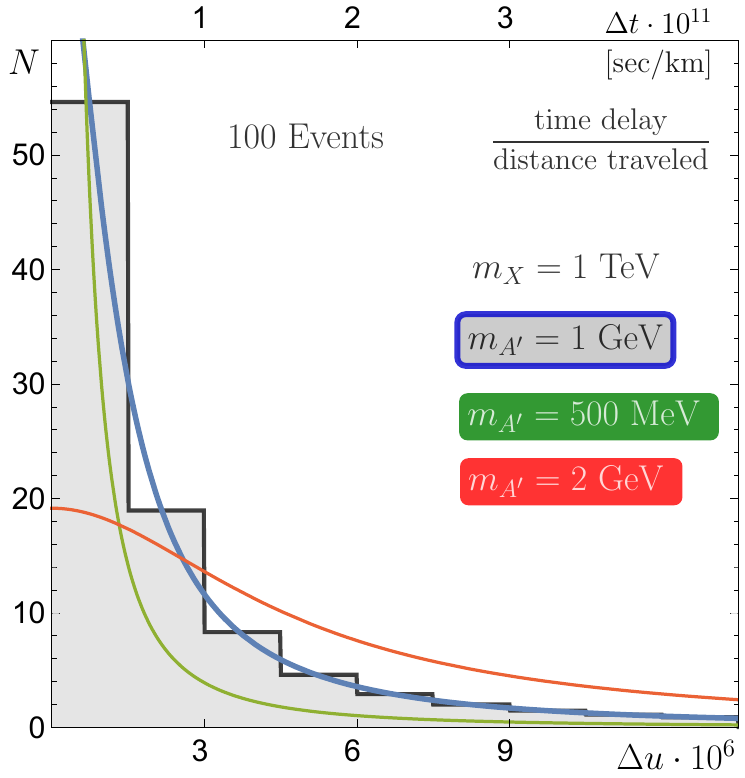} 
\includegraphics[width=0.32\linewidth]{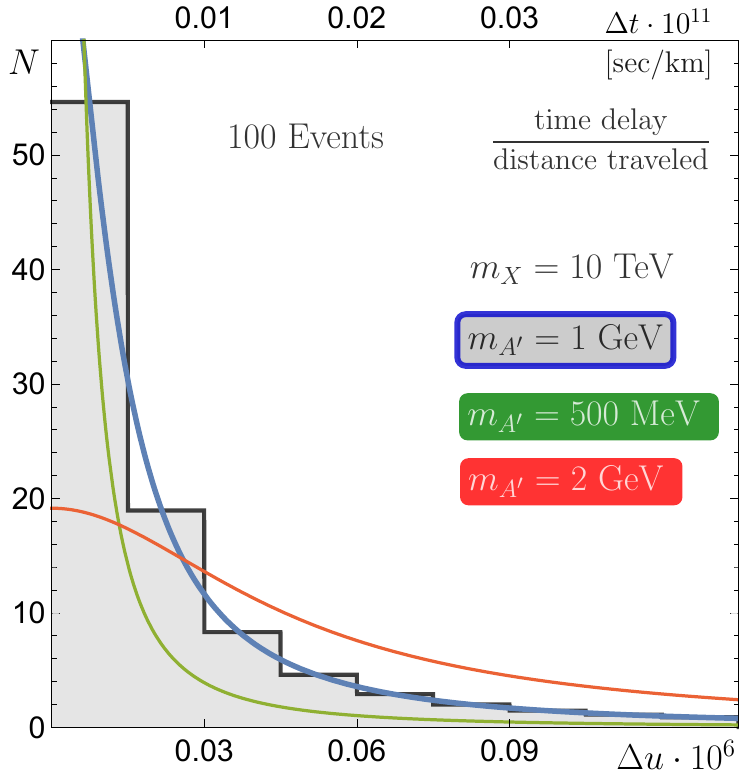} 
\vspace*{-0.1in} 
  \caption{%
	Lab-frame muon velocity differences for
	$A'\to \mu^+\mu^-$ and
    $(m_X,m_{A'}) = (100~\gev, 500~\mev)$
    (left), $(1~\tev, 1~\gev)$ (center), and $(10~\tev, 1~\gev)$
    (right).  Distributions are normalized to 100 events and
    different values of $m_{A'}$ are shown for
    comparison.  The top axes measure the time
    delay between the two final states per km
    between the decay and observation positions. 
  \label{fig:results:velocity}}
\vspace*{-0.1in}
\end{figure}

The dark photon signal has two primaries, which could in principle be
identified as a smoking-gun signal for the dark sector.  In
\figsref{results:velocity}{results:opening angle} we show histograms
of the velocity difference (time delay) and opening angle (track
separation) of the two muons produced in a dark photon decay.
Details of the distributions are presented in the Appendix.
Parallel tracks have been
considered previously in the context of slepton production from high
energy neutrinos in Refs.~\cite{Albuquerque:2003mi,Albuquerque:2006am}
and have recently been searched for by
IceCube~\cite{Kopper:2015:ICRC}.  As a benchmark for IceCube reach,
the parameters $m_X = 1~\tev$, $m_{A'} = 500~\mev$, and $\varepsilon =
5 \times 10^{-9}$ gives an expected 40 muon events in 10 live
years. The center panels of \figsref{results:velocity}{results:opening
  angle} then show that over $\sim 2.5~\km$ between the $A'$ decay
point and the maximal detection distance, one expects a few events
with timing separation of $\sim 0.03$~ns and $\sim 20$~m track
separation.  

The timing separation is below the IceCube Digital
Optical Module timing resolution of $\sim 5$~ns~\cite{Halzen:2010yj}.
The track separations are less than the $\sim 100~\m$ separations
probed by current analyses~\cite{Kopper:2015:ICRC}, but they are also
much larger than the $\sim 1~\m$ separations from SM neutrino-induced
charm production.  These results motivate looking for parallel muon
tracks with ${\cal O}(10~\m)$ separations, which would be an
unambiguous signal of physics beyond the SM, and a spectacular signal
of dark photons and dark matter.  
One step in this direction is the proposed PINGU upgrade which would densely instrument a subset of the IceCube/DeepCore detector~\cite{Aartsen:2014oha}. However, for the proposed dark photon search, this comes at a large cost in available volume and propagation distance.
A possible alternative direction to improve sensitivity to these parallel muon signals is to increase the detector density of the IceTop surface array.

\begin{figure}[t] 
\includegraphics[width=0.32\linewidth]{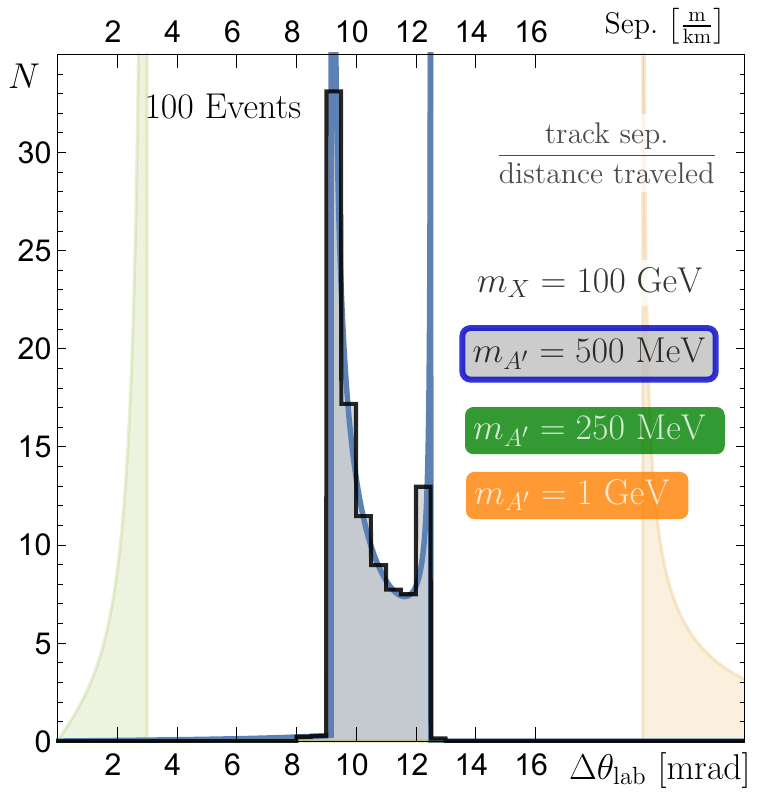}
\includegraphics[width=0.32\linewidth]{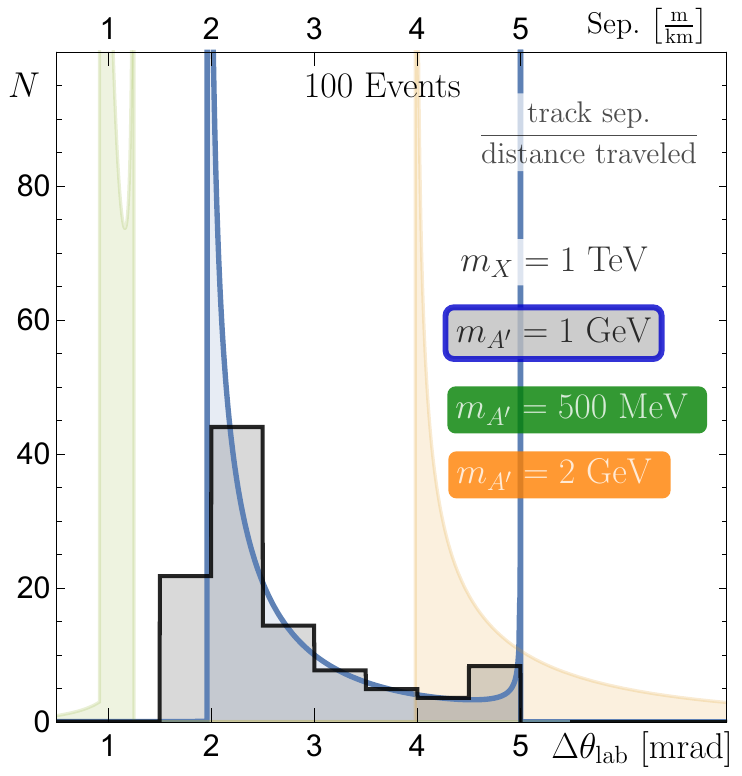}
\includegraphics[width=0.32\linewidth]{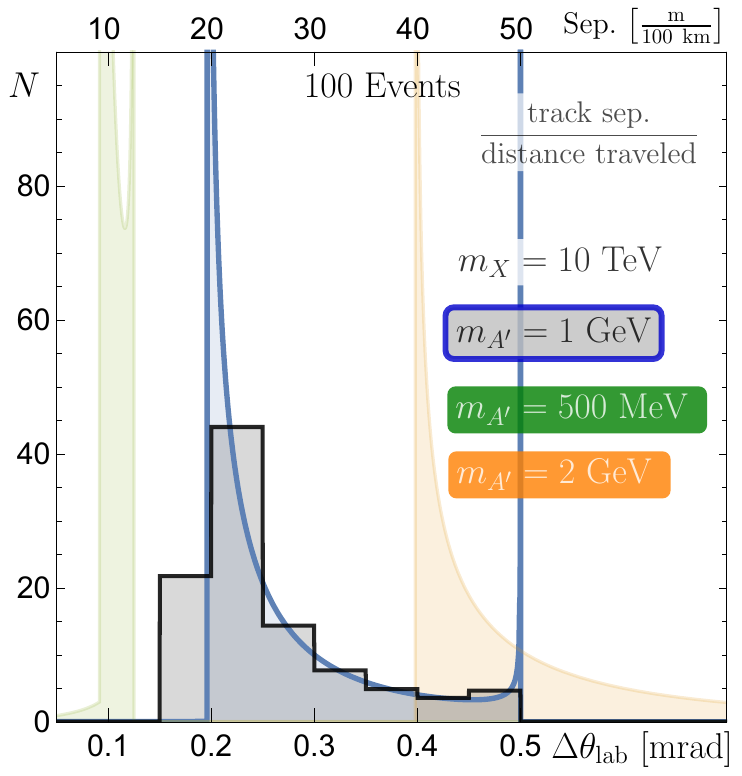} 
\vspace*{-0.1in} 
\caption{Lab-frame muon opening angles (bottom axis)/track
  separation (top axis) for $A'\to\mu^+\mu^-$ and the same $m_X$ values as 
  \figref{results:velocity}.  Different values of
  $m_{A'}$ are shown for comparison. 
  } 
  \label{fig:results:opening angle}
\vspace*{-0.1in}
\end{figure}

\subsection{Fermi-LAT/AMS-02}

The dark photon decay products may also be detected by space-based
cosmic ray detectors, such as Fermi-LAT and AMS-02.  Though these are
far smaller than IceCube, the dark photon may decay anywhere between
the Earth's surface and the detector, providing a
partially-compensating enhancement to the rate.  For Fermi and AMS, we
follow the formalism described above, but now use $A_{\text{eff}} =
1~\m^2$.  Both Fermi and AMS are in low Earth orbit, flying 550 and
400 km above the ground, respectively.  We choose $D = 550~\km$ in
\eqref{eq:decay:efficiency}.  Note that, after the dark photon decays,
the resulting charged particles are bent in the Earth's magnetic field
by an angle
\begin{equation}
\theta = 0.5^{\circ} \, \frac{\tev}{p} \, \frac{L}{550~\km} \ 
\frac{B}{0.5~\text{G}} \ ,
\label{eq:B:deflection}
\end{equation}
where $p$ is the particle's momentum, $L$ is the distance it travels,
and we have normalized the Earth's magnetic field $B$ to an average
value at the surface of the Earth.  For $m_X \agt \tev$, this
deflection is less than the dispersion from the dark matter's spatial
distribution at the center of the Earth given in
\eqref{eq:DM:radius:earth}, but for $m_X \sim 100~\gev$, this
deflection may be significant, and the signal may arrive at an angle
as large as $5^{\circ}$ relative to straight down.

\begin{figure}[t] 
\includegraphics[width=0.32\linewidth]{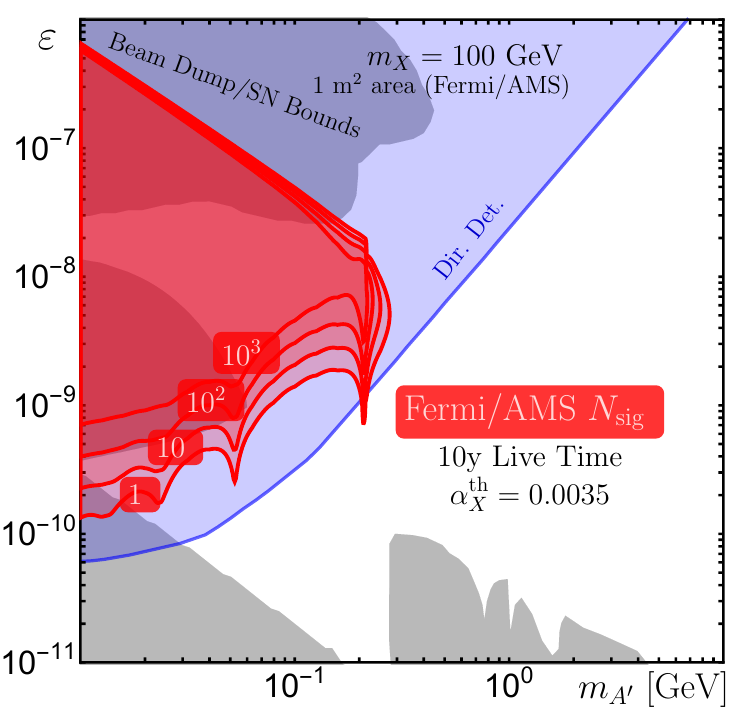}
\includegraphics[width=0.32\linewidth]{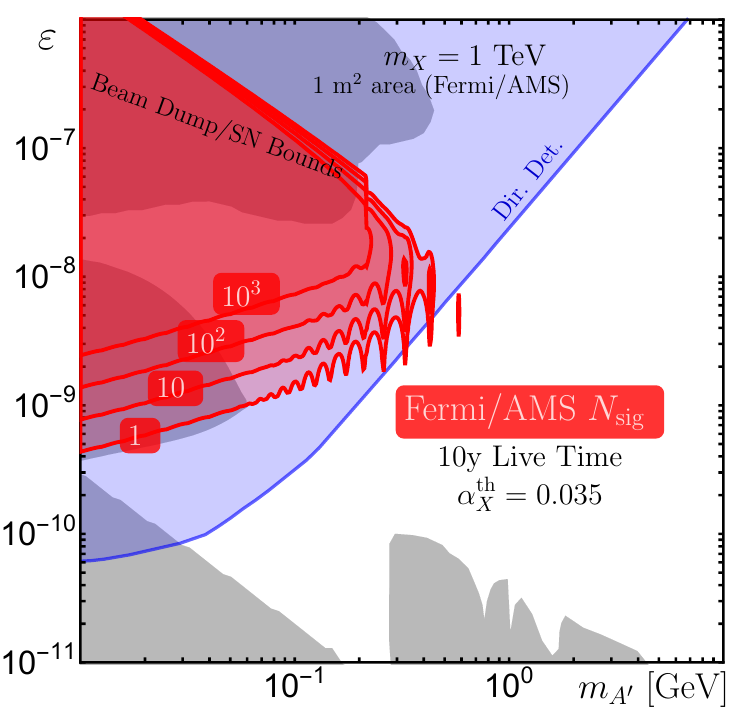}
\includegraphics[width=0.32\linewidth]{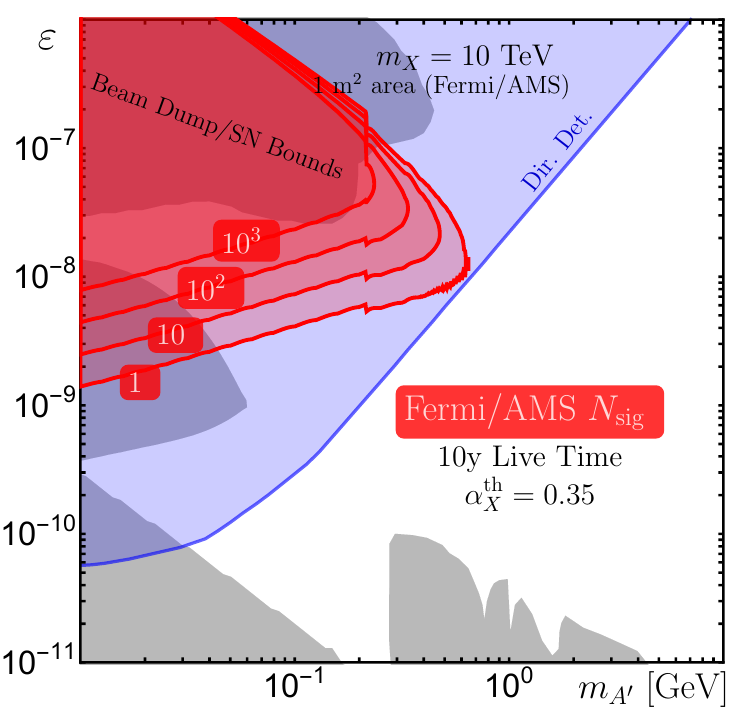}
\vspace*{-0.1in} 
  \caption{10 year signal event rates for a space-based detector in
  low Earth orbit, such as Fermi/AMS. Colors and bounds are the same as 
  \figref{resultsIceCube}.}
  \label{fig:resultsFermiAMS}
\vspace*{-0.1in}
\end{figure}

The resulting event rates for such space-based detectors are given in
\figref{resultsFermiAMS} for a live time of 10 years. 
The parameter space that can be probed largely overlaps
with that already probed by direct detection, but Fermi and AMS may set
bounds complementary to the existing direct detection experiments. 
As a benchmark, consider the
parameters $m_X = 1~\tev$, $m_{A'}=400~\mev$, and $\varepsilon=10^{-8}$, for which one expects $N_\text{sig} = 10$ signal
events in 10 live years. The velocity difference and opening angle
distributions are shown in the center panels of
\figsref{results:velocity}{results:opening angle}. For a primary
propagation distance of $\sim 300~\km$, this yields timing separations
of up to tens of nanoseconds and separations of up to a kilometer.  We
therefore do not expect to see both primary particles from dark photon
decay in Fermi or AMS.  Of course, this is still possible: 
the $A'$ may decay near Fermi or AMS; 
secondary photons from hadronic final states my happen to have
little transverse momentum;
or the $A'$ may decay far from the
detectors to two charged particles, which are both bent by the
magnetic field into the detectors.  
Although possible, all of these
are highly improbable, and two-particle events are a small fraction of
the total number of single-particle signal events. An alternative possibility
is when there is a small splitting between $m_{A'}$ and $2m_f$. In this case the decay products have small transverse momentum by~\eqref{eq:opening:angle}, at the cost of a reduced branching ratio.

Last, the number of signal events $N_{\text{sig}}$ does not take into
account experimental efficiencies associated with each apparatus. For
example, we have assumed that the volume of the International Space
Station between the Earth and AMS does not affect the dark photon
primaries, and, further, that the hadronic products of the dark photons
are detectable. A more complete analysis of the Fermi/AMS reach will
require more realistic modeling and different triggers.

\section{Conclusions} 

We have presented a novel method to discover dark matter that
interacts with the known particles through dark photons that
kinetically mix with the SM photon. The dark matter is captured by the
Earth and thermalized in the Earth's center, and then annihilates to
dark photons.  The dark photons then travel to near the surface of the
Earth and decay. We have determined the signal rates without
simplifying assumptions concerning the dark matter and dark photon
masses.  In viable regions of the model parameter space, thousands of
such dark photon decays are possible in IceCube, and smaller, but
still detectable, signals in space-based detectors such as Fermi and
AMS are also possible.

As with traditional indirect detection signals that rely on
annihilation to neutrinos, the dark photon signal points back to the
center of the Earth, differentiating it from astrophysical
backgrounds.  In contrast to the neutrino signal, however, the dark
photon decays to two visible particles.  The dark photon signal is
therefore even more striking, as it is monoenergetic if fully
contained.  In addition, in principle both particles could be detected
simultaneously yielding, for example, parallel muon tracks in IceCube
with separations of $\sim {\cal O}(10~\m)$.  We have shown
distributions of these separations for representative points in model
parameter space.

As discussed in \secref{capture}, the leading uncertainty in the
signal rate predictions is from the capture rate analysis.  The escape
velocity of the Earth is not large, and so this capture rate is
subject to detailed modeling, including the effects of the Earth, Sun,
Jupiter, and Venus.  In addition, a cold ``dark disk'' population of
dark matter may significantly enhance the capture rates.  The
implications of these effects for WIMP dark matter have been
considered in Refs.~\cite{Lundberg:2004dn,Peter:2009mi,%
  Peter:2009mk,Peter:2009mm,Bruch:2009rp,Purcell:2009yp}; it would be
interesting to determine their effects on dark matter with dark
photon-mediated interactions.

In this study, we have assumed the dark matter $X$ is a Dirac fermion
and the mediator is a dark photon that mixes only with the SM photon,
and so couples only to charged particles.  It would be interesting to
consider cases where $X$ is a pseudo-Dirac fermion or a scalar, and
cases where the dark photon mixes with the $Z$ (and so couples to
neutrinos, for example), or is replaced by a scalar (for which the
dark matter may also be Majorana).  Dark matter that collects and
annihilates at the center of the Sun is also a promising source of
decaying dark photons and will probe different regions of parameter
space~\cite{Feng:2016ijc}.

Finally, the experiments have been modeled very roughly here; detailed
analyses, preferably by the experimental collaborations themselves,
are required to evaluate the accuracy of the signal rate estimates.
However, our conclusion that there are viable regions of parameter
space that predict thousands of signal events indicates that there are
certainly regions of parameter space where the indirect detection
signals discussed here are the most sensitive probes, surpassing
direct detection detectors, beam dump experiments, and cosmological
probes.  The possibility of discovering signals of dark matter that,
unlike so many other indirect detection signals, are essentially free
of difficult-to-quantify astrophysical backgrounds, provides a strong
motivation for these searches.

\section{Acknowledgments} 

\noindent
We thank Ivone Albuquerque, James Bullock, Gustavo Burdman, Eugenio
Del Nobile, Francis Halzen, Simona Murgia, Maxim Pospelov, Brian
Shuve, Tim M.P.~Tait and Hai-Bo Yu for helpful discussions. We thank Adam Green for pointing out a typo in our decay length code, which moved the region of experimental sensitivity to values of $\varepsilon$ that are lower by an order of magnitude. The work of J.L.F.\ and P.T.\ was performed in part at the Aspen Center 
for Physics, which is
supported by National Science Foundation grant PHY--1066293.
P.T.~thanks the Munich Institute for Astro- and Particle Physics
(MIAPP, DFG cluster of excellence "Origin and Structure of the
Universe") workshop ``Anticipating Discoveries: LHC14 and Beyond'' for
its hospitality and support during the part of this work.
J.S.~and~P.T.\ thank UC Davis and the (Pre-)SUSY 2015 conference for
its hospitality during the completion of this work.  This work is
supported in part by NSF Grant No.~PHY--1316792.  J.L.F.\ was
supported in part by a Guggenheim Foundation grant and in part by
Simons Investigator Award \#376204.  P.T.~is supported in part by a
UCI Chancellor's ADVANCE fellowship.  Numerical calculations were
performed using \emph{Mathematica 10.2}~\cite{Mathematica10}.

\appendix*

\section{Decay Product Distributions}
\label{app:deca:distributions}

We summarize analytic results for the kinematic distributions of the
dark photon decay products, presenting the forward velocity difference
between the two final states and the lab frame opening angle, which
may be used to determine the time delay and track separation between
these objects in a detector. 
For simplicity, we assume the dark photon
decays isotropically in its rest frame.  Angular correlations will
modify our distributions, but will not change the ranges of time delay
and track separation, which are our primary interest.  With this
approximation, 
in the center-of-mass frame, the dark photon decay products are
evenly distributed in $\cos \theta_{\text{CM}}$, 
where $\theta_\text{CM}$ is the
angle between the dark photon boost direction and one of the decay
products.  The value of a kinematic quantity $k$ for fixed model
parameters is a function $\kappa$ of $\cos \theta_{\text{CM}}$. The
distribution of these values $f$ is
\begin{align}
	f(k) 
	&= 
	\sum_{\cos\theta_\text{CM}^i}^{\kappa(\cos\theta_\text{CM}^i)=k} 
	\frac{1}{\left| \kappa'(\cos\theta_\text{CM}^i) \right|}
	\ .
\end{align}

Throughout this appendix we consider two-body decays $A' \to f
\bar{f}$ and define
\begin{align}
	a &= \frac{2m_f}{m_{A'}}
	&
	b &= \frac{m_{A'}}{m_X} \ .
	\label{eq:app:a:b:c:def}
\end{align}

\subsection{Velocity Distribution and Time Delay}

\begin{figure}[t] 
\begin{center}
\includegraphics[width=0.4\linewidth]{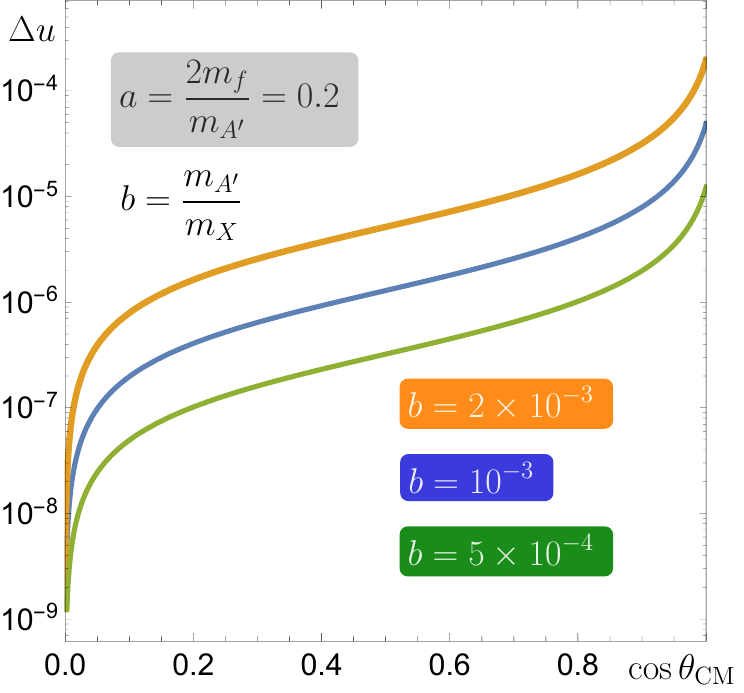}
\qquad
\includegraphics[width=0.4\linewidth]{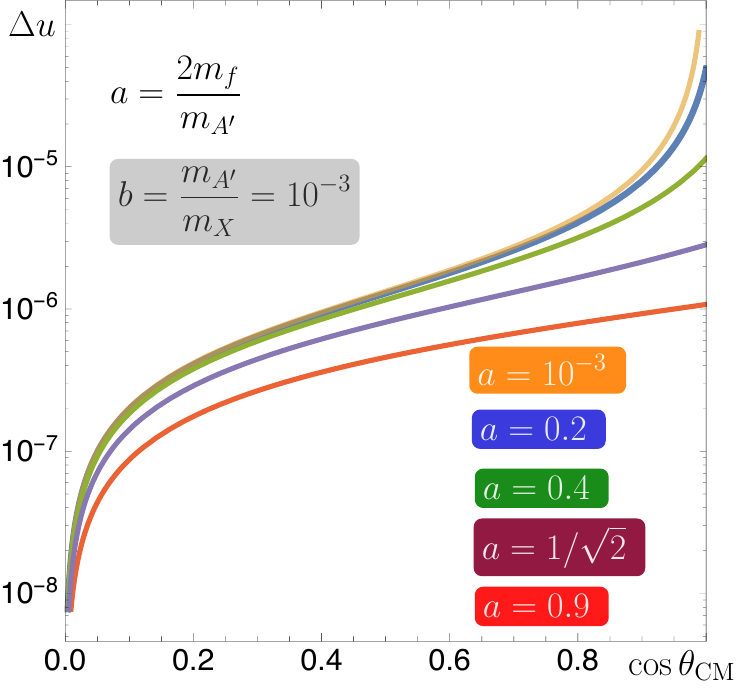}
\end{center}
\vspace*{-0.1in} 
  \caption{%
  Velocity difference, $\Delta u$, of the two particles produced in $A'$ decay as a function of the center-of-mass frame angle $\cos\theta_\text{CM}$ for representative values of $a$ and $b$.
  }
  \label{fig:app:delta:u:only}
\vspace*{-0.1in}
\end{figure}

In the Earth's rest frame, the forward velocities of the particles
produced in $A'$ decay are
\begin{align}
	u_\pm &= \frac{
	\sqrt{1-b^2}\pm \sqrt{1-a^2} \cos\theta_{\text{CM}}}
        {1 \pm \sqrt{1-b^2} \sqrt{1-a^2} \cos\theta_{\text{CM}} } \ ,
	\end{align}
where we use natural units $c=1$.  The difference of these velocities
is
\begin{equation}
\Delta u \equiv u_+ - u_- 
= \frac{2 b^2 \sqrt{1-a^2} \cos\theta_{\text{CM}}}{1-(1-b^2)(1-a^2) 
\cos^2\theta_{\text{CM}} } 
\approx 
\frac{2 b^2 \sqrt{1-a^2} \cos\theta_{\text{CM}}}{1-(1-a^2) 
\cos^2\theta_{\text{CM}} } \ ,
	\label{eq:Delta:u:from:c}
\end{equation}
where the last expression is valid for $b \ll 1$, the values we are
most interested in. We plot $\Delta u(\cos\theta_\text{CM})$ in \figref{app:delta:u:only}. Observe that $\Delta u$ scales like $b^2$ for small $b$; this is also seen in \figref{results:velocity}, where the $m_X=1~\tev$ and $m_X=10~\tev$ plots are related by a simple rescaling. Further, the
distribution is fairly insensitive to $a=2m_\ell/m_{A'}$ for $m_{A'}
\sim$~GeV and for $\ell=e,\mu$.

For a given $\Delta u$, the (dimensionful) time delay between the two
decay products for a decay that occurs a distance $L$ from the
detection point is
\begin{align}
	\Delta t &= \frac{L}{cu_-} - \frac{L}{cu_+}
	=
	\frac{L \Delta u}{c u_- u_+}
	\approx \frac{L}{c} \Delta u \ ,
\end{align}
where we've taken the limit of large boost so that $u_{\pm} \to 1$.

\subsection{Opening Angle and Track Separation}

In the Earth's rest frame, the angles $\theta_\pm$ of the decay
products relative to the $A'$ decay direction are
\begin{align}
	\tan \theta_\pm 
	&=
	\frac{\pm b \sqrt{1-a^2} \sin\theta_{\text{CM}}}
        {\sqrt{1-b^2}\pm \sqrt{1-a^2} \cos\theta_{\text{CM}} } \ .
	\label{eq:app:tan:theta:pm}
\end{align}
The opening angle between the two decay products
is therefore
\begin{equation}
\Delta\theta_\text{lab} \equiv \tan^{-1}\theta_+ - \tan^{-1}\theta_- 
\approx \frac{2 b \sqrt{1-a^2} \sin\theta_{\text{CM}} }
{1 - (1-a^2) \cos^2 \theta_{\text{CM}}} \ ,
\end{equation}
where the last expression is valid for $b \ll 1$. 
The scaling $\Delta\theta_\text{lab}\propto b$ can be seen in the center
and right plots in \figref{results:opening angle}. 
The maximal opening angle is
\begin{equation}
\Delta\theta_{\text{lab}}^{\text{max}}
= \left\{ \begin{array}{ll}
2 b \sqrt{1-a^2} \ \text{at} \ \cos \theta_{\text{CM}} = 0 \ , 
\quad & \displaystyle a \ge \frac{1}{\sqrt{2}} \\
\displaystyle\frac{b}{a} \ \text{at} \ \cos \theta_{\text{CM}} 
= \sqrt{\frac{1-2a^2}{1-a^2}} \ ,  
\quad & \displaystyle a < \frac{1}{\sqrt{2}}  
\end{array} \right. \ . 
\end{equation}

\begin{figure}[t] 
\begin{center}
\includegraphics[width=0.4\linewidth]{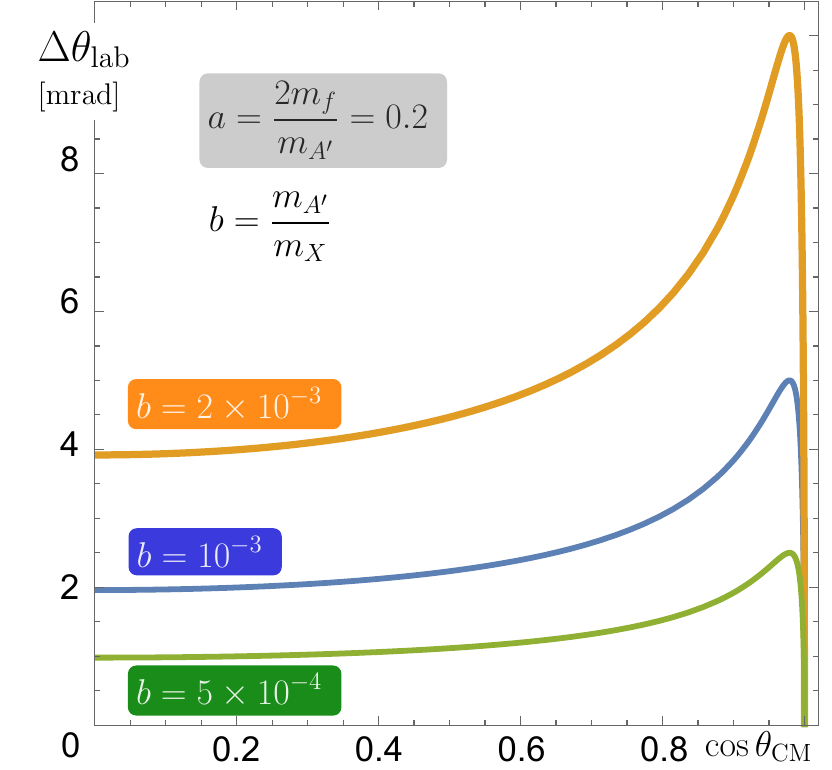}
\qquad
\includegraphics[width=0.4\linewidth]{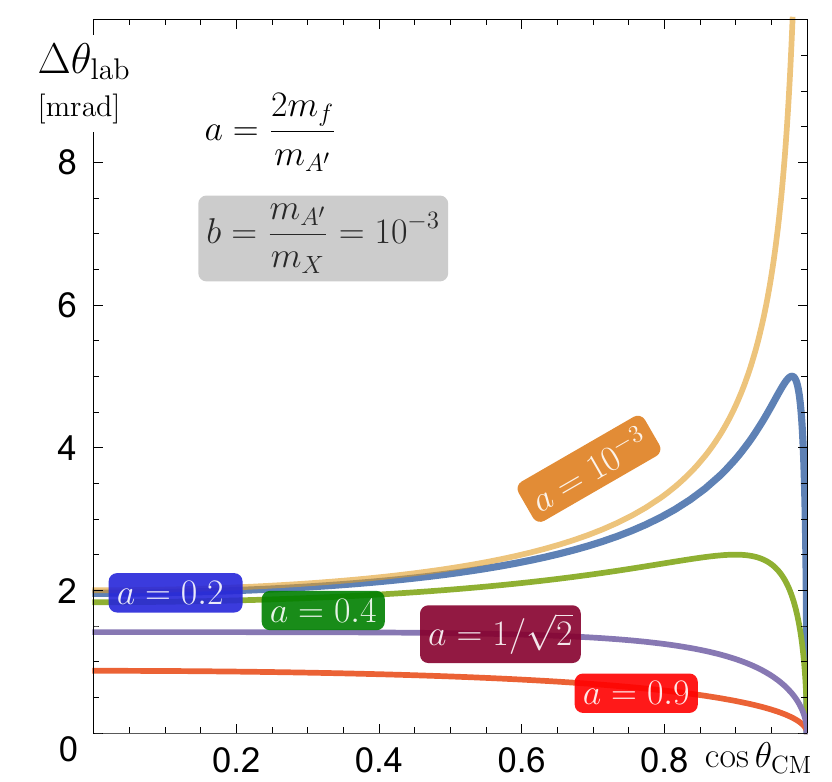}
\end{center}
\vspace*{-0.1in} 
  \caption{Earth-frame opening angle between the two 
  particles produced in $A'$ decay as a function of the center-of-mass
    frame angle $\cos\theta_\text{CM}$ for representative values of $a$ and $b$.
}
  \label{fig:app:delta:theta:only}
\vspace*{-0.1in}
\end{figure}

We plot $\Delta\theta_\text{lab}(\cos\theta_{\text{CM}})$ in \figref{app:delta:theta:only}.  
For large $a$, the opening angle is
maximized at $\cos\theta_{\text{CM}} = 0$, consistent with the
intuition that the largest opening angle should correspond to fully
transverse decays in the center-of-mass frame.  But for small $a$,
this intuition does not hold: the maximal opening angle occurs for
$\cos\theta_{\text{CM}} \approx 1$, where one particle is emitted
``backwards'' in the $A'$ center-of-mass frame so that its forward
velocity is significantly reduced, enlarging the opening angle.  In
most of the range of $\cos\theta_{\text{CM}}$, 
$\Delta\theta_\text{lab} \approx 2b$, but
the maximal opening angle $\Delta\theta_\text{lab}^{\text{max}}
\approx b/a$ occurs for large $\cos\theta_{\text{CM}} \approx 1 -
\frac{1}{2} a^2$.

\begin{figure}[t] 
\begin{center}
\includegraphics[width=0.4\linewidth]{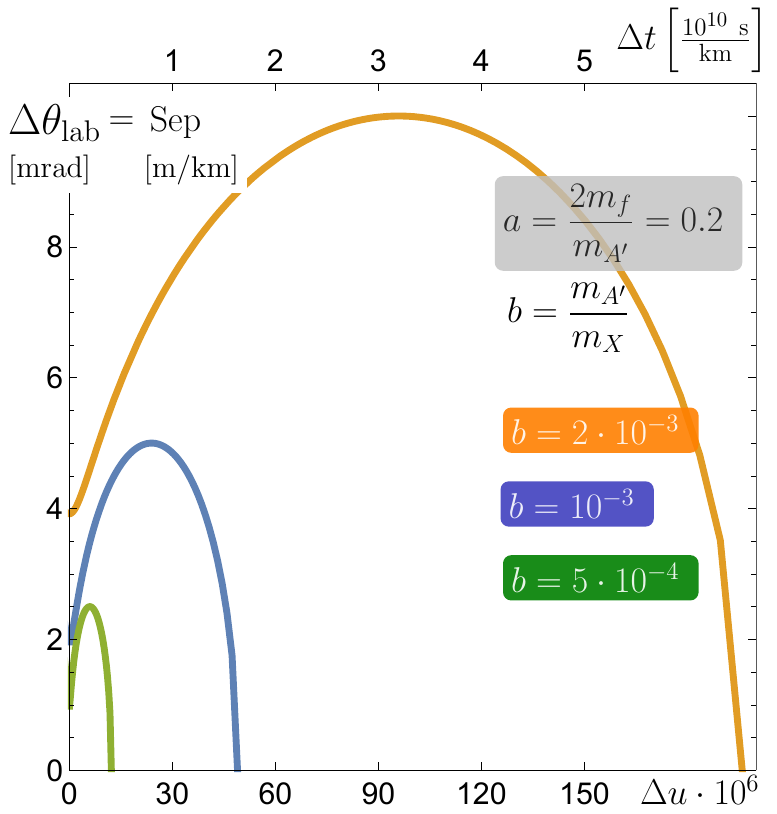} 
\qquad
\includegraphics[width=0.4\linewidth]{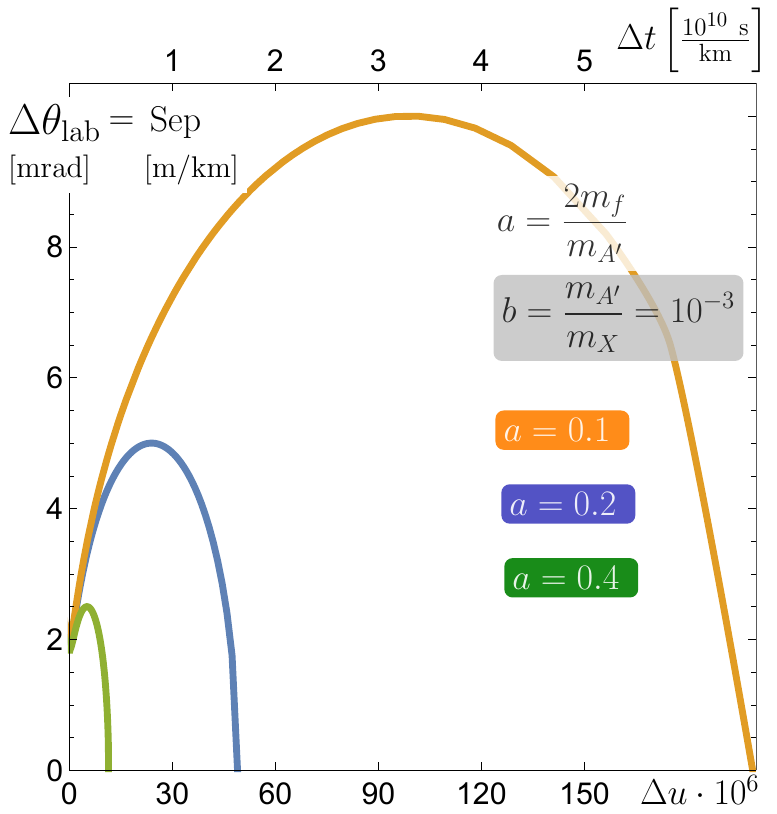}
\end{center}
\vspace*{-0.1in} 
  \caption{
    The correlation between the velocity difference and the lab-frame opening angle  of the two particles produced in $A'$ decay for representative values of $a$ and $b$.
}
  \label{fig:app:correlation:only}
\vspace*{-0.1in}
\end{figure}

Finally we show the correlation between $\Delta\theta_\text{lab}$ and
$\Delta u$ in \figref{app:correlation:only}.  These plots identify where one may use the combination of the opening angle and time delay to discriminate the two final state particles.

\bibliography{bibdarkphotonearth}

\end{document}